\documentclass[fleqn,usenatbib]{mnras}

\usepackage{newtxtext,newtxmath}
\usepackage[T1]{fontenc}
\usepackage{ae,aecompl}

\usepackage{graphicx}
\usepackage{bm}		
\usepackage[flushleft]{threeparttable}
\usepackage{float}
\usepackage[caption = false]{subfig}
\usepackage{hyperref}
\AtBeginShipout{%
  \ifnum\value{page}>1 %
    \typeout{* Additional boxing of page `\thepage'}%
    \setbox\AtBeginShipoutBox=\hbox{\copy\AtBeginShipoutBox}%
  \fi}

\usepackage{multicol}        

\usepackage{pdflscape}
\usepackage{adjustbox}
\usepackage{threeparttable}

\title[The nature of X-ray spectral variability in MCG--6-30-15]{The nature of X-ray spectral variability in MCG--6-30-15}

\author[E. S. Kammoun and I. E. Papadakis]{E. S. Kammoun,$^{1}$\thanks{E-mail: \href{mailto:ekammoun@sissa.it}{ekammoun@sissa.it}}
and I. E. Papadakis$^{2,3}$ 
\\
$^{1}$SISSA, via Bonomea 265, I-34135 Trieste, Italy\\
$^{2}$Department of Physics and Institute of Theoretical and Computational Physics, University of Crete, 71003, Heraklion, Greece\\
$^{3}$IESL, Foundation of Research and Technology, 71110 Heraklion, Greece\\
}
\date{Accepted XXX. Received YYY; in original form ZZZ}

\pubyear{2017}

\begin{document}
\label{firstpage}
\pagerange{\pageref{firstpage}--\pageref{lastpage}}

\maketitle

\begin{abstract}
{The flux-flux plot (FFP) method can provide model-independent clues regarding the X-ray variability  of active galactic nuclei. To use it properly, the bin size of the light curves should be as short as possible, provided the average counts in the light curve bins are larger than $\sim 200$.  We apply the FFP method to the 2013, simultaneous {\it XMM-Newton} and {\it NuSTAR} observations of the Seyfert galaxy MCG--6-30-15, in the 0.3--40~keV range. The FFPs above $\sim 1.6$ keV are well-described by a straight line. This result rules out spectral slope variations and the hypothesis of absorption driven variability. Our results are fully consistent with a power-law component varying in normalization only, with a spectral slope of $\sim 2$,  plus a variable, relativistic reflection arising from the inner accretion disc around a rotating black hole. We also detect spectral components which remain constant over $\sim 4.5$\,days (at least). At energies above $\sim 1.5$ keV, the stable component is consistent with reflection from distant, neutral material. The constant component at low energies is consistent with a blackbody spectrum of $kT_{\rm BB} \sim 100$\,eV. The fluxes of these components are $\sim 10-20\%$ of the average continuum flux (in the respective bands). They should always be included in the models that are used to fit the spectrum of the source. The FFPs below 1.6\,keV are non-linear, which could be due to the variable warm absorber in this source.}

\end{abstract}



\section{Introduction}
\label{sec:intro}

According to the current paradigm,  active galactic nuclei (AGN) 
are thought to be powered by accretion of matter, in the form of a disc, onto a central 
supermassive black hole (BH) of mass $M_{\rm BH} \sim 10^{6-9}\,M_{\odot}$. 
AGN are strong X-ray emitters, and it is widely accepted that the X--rays are produced by Compton up-scattering of ultraviolet (UV)/soft X-ray disc photons off hot electron \citep[$\sim 10^9$\,K; e.g.][]{Shap76, Haa93}. Since the X--ray luminosity is a substantial part of the bolometric luminosity in these objects, it is believed that the X--ray source (which is usually referred to as the `X--ray corona') is located close to the central black hole, where most of the accretion power is released. AGN are highly variable in X--rays, both in flux and spectral shape. The amplitude of the X--ray flux variations is the highest, and the variability time scales are the shortest,  among the variations at all wavelengths in radio-quiet AGN. This observational characteristic indicates that the X--ray source should also be small. Because of these  characteristics, it is believed that X--ray spectral and timing studies can provide important clues regarding the physical processes that operate in the innermost region of AGN. 

In this work, we apply the flux--flux plot (FFP) method to the simultaneous {\it XMM-Newton} and {\it NuSTAR} observations of the Seyfert 1 galaxy MCG--6-30-15 ($z = 0.00775$), performed in January 2013. Our main objective is to study its X--ray flux and spectral variability properties.  The FFP method was first developed by \cite{Chu01} and was applied to the study of the X--ray variability of the black hole binary Cygnus X-1. It was first applied to AGN studies by \cite{Tay03}, with the aim to study the X-ray spectral variability of X-ray bright Seyferts. It has been used since then in numerous AGN X--ray variability  studies. 

MCG--6-30-15 is the archetype of Seyferts with broad iron lines in their X--ray spectra. It was the first source where a broad Fe K$\alpha$ line with a red tail was detected. The line shape was interpreted as being due to relativistic reflection, implying an almost maximally spinning Kerr black hole \citep[e.g.][]{Tana95, Iwa96, Iwa99, Min07, Mari14}. This interpretation was supported by the detection of short delays between the X--ray continuum and the soft band (i.e. X--rays below $\sim 1.5$ keV) emission \citep[e.g.][]{Emmanou11, Emma14, Kara14}. \cite{Epitropakis16} showed that the iron line/continuum time delays are consistent with the delays between the hard (i.e. $>2$ keV) and soft band variations.

MCG-6-30-15 is highly variable in X--rays. It shows large amplitude flux and spectral variations on short (minutes/hours) and long (days/years) time scales. Its spectral variations have been interpreted within the context of a two component model which consists of: 1) a 
highly variable power-law (PL) continuum (with an almost constant spectral 
slope of $\Gamma \sim 2$), and 2) a less variable ionized reflection spectrum 
arising within a few gravitational radii \citep{Shih02, Fab03, Tay03, Parker14}.  The soft X-ray spectrum of the source is affected by a complex 
warm absorber \citep[e.g.][]{Otani96, Reynolds97, Brand01,Turner03, Turner04}, whose properties vary in time, and should add to the observed variability of the source. In fact, \cite{Mil08, Mil09} proposed a complex absorption-dominated model 
in order to explain the red-tail of the iron line and the spectral variability of MCG--6-30-15. According to this model, partial-covering absorbers in the line of sight (having column 
densities in the $10^{22}-10^{24}\,{\rm cm^{-2}}$ range), can  
produce an apparent broadening of the Fe K$\alpha$ line similar to the one 
caused by relativistic effects \citep[e.g.][]{Mil07,Turner07}. Variability in the covering fraction of these absorbers could also explain the observed spectral variations.

We recently applied the FFP method to the narrow-line Seyfert 1 galaxy IRAS\,13224--3809 \citep{Kam15}. We found that, if the source is highly variable and the intrinsic FFP is non-linear, the shape of the observed FFPs may be affected by the light curves' bin size. We suggested the use of the shortest possible bin size in the construction of the FFPs. In this work, we investigate the effects of the Poisson noise to the observed FFPs, and we provide practical guidelines for their estimation. 

Although the main objective in the past applications of the FFP method was the determination of constant spectral components, in this work we use the FFPs to also study the 
variable spectral components in the X--ray spectrum of MCG-6-30-15. FFPs can provide model-independent information on the origin of the spectral variability in AGN, and MCG--6-30-15 is an ideal target for this: it is highly variable and X--ray bright. As a result, we can use the {\it NuSTAR} data to study the FFPs at energies up to 40 keV.  This was not possible to achieve in the case of IRAS\,13224--3809, whose flux is low above $\sim 3-4$ keV. We detect a constant component at energies above $\sim 1.5$ keV, which is indicative of X-ray reflection from neutral material. We find that the hard X-ray emission is variable in amplitude, but not in shape (contrary to IRAS\,13224--3809), and that it cannot be due to absorption related variations only. Similar to IRAS\,13224--3809, we find strong evidence of a variable X--ray reflection component originating from an ionized disc, which extends to the inner stable circular orbit around a maximally rotating BH. We also find evidence of a constant component at low energies, which may arise from the inner disc. 

\section{Observations and data reduction}
\label{sec:obsred}

\subsection{\it XMM-Newton}
\label{subsec:XMMdata}

The {\it XMM-Newton} sattelite \citep{Jans01} observed MCG--6-30-15 
simultaneously with {\it NuSTAR} \citep{Har13}, starting on 
2013 January 29 during three consecutive revolutions 
(Obs. IDs 0693781201, 0693781301, and 0693781401). The data are available 
in the {\it XMM-Newton} Science Archive\footnote{\url{http://nxsa.esac.esa.int/nxsa-web}} 
(XSA). We considered data provided by the EPIC-pn camera 
\citep{Stru01} only, that was operating in small window/medium 
filter imaging mode. 
{We do not consider the data from the two EPIC-MOS \citep{Tur01} detectors because they were affected by a high level of pile-up \citep{Mari14}.}

We reduced the data using the {\it XMM-Newton} Science Analysis System 
({\tt SAS}\,v15.0.1) and the latest calibration files. The data were 
cleaned for strong background flares and were selected using the criterion 
PATTERN\,$\leq$\,4. Source light curves were extracted from a circle of radius 
40\arcsec, while the background light curves were extracted from an off-source 
circular region of radius 50\arcsec. We checked for pileup and we found it 
to be negligible in all observations. Background-subtracted 
light curves were produced using the {\tt SAS} task {\tt EPICLCCORR}.

\subsection{\it NuSTAR}
\label{subsec:Nustardata}

MCG--6-30-15 was observed by {\it NuSTAR} with its two co-aligned
telescopes with corresponding Focal Plane Modules\,A (FPMA)
and B (FPMB) starting on 2013 January 29 (Obs. IDs 60001047002, 60001047003, 
and 60001047005). We reduced the {\it NuSTAR} data following the standard 
pipeline in the {\it NuSTAR} Data Analysis Software (NuSTARDAS\,v1.6.0).
We used the instrumental responses from the latest calibration files available
in the {\it NuSTAR} calibration database (CALDB). The unfiltered event files 
were cleaned with the standard depth correction, which reduces the internal 
background at high energies, and we excluded South Atlantic Anomaly passages 
from our analysis. The source and background light curves were extracted from 
circular regions of radii 1\farcm5 and 3\arcmin, respectively, for both FPMA 
and FPMB, using the HEASoft task {\tt NUPRODUCT}, and requiring an exposure 
fraction larger than 50\,\%. We checked that the background-subtracted light 
curves of the two {\it NuSTAR} modules were consistent with each other as 
follows. We divided the FPMA over the FPMB light curves (binned at 
$\Delta t = 1\,{\rm ks}$), in all the energy bands we consider in this work (see next Section), and we fitted the ratio as a function of time with 
a constant, $C$. The fit was acceptable in all cases,  
indicating that the FPMA and FPMB light curves are consistent ($C$ being consistent with 1 in all cases). Given this result, we added the FPMA and FPMB light curves in the various energy bands considered in this work, using the {\tt FTOOLS} \citep{Ftools} command 
{\tt LCMATH}, in order to increase the signal-to-noise of the {\it NuSTAR} light curves. 

\begin{figure*}
\centering
\includegraphics[width =1.0\textwidth]{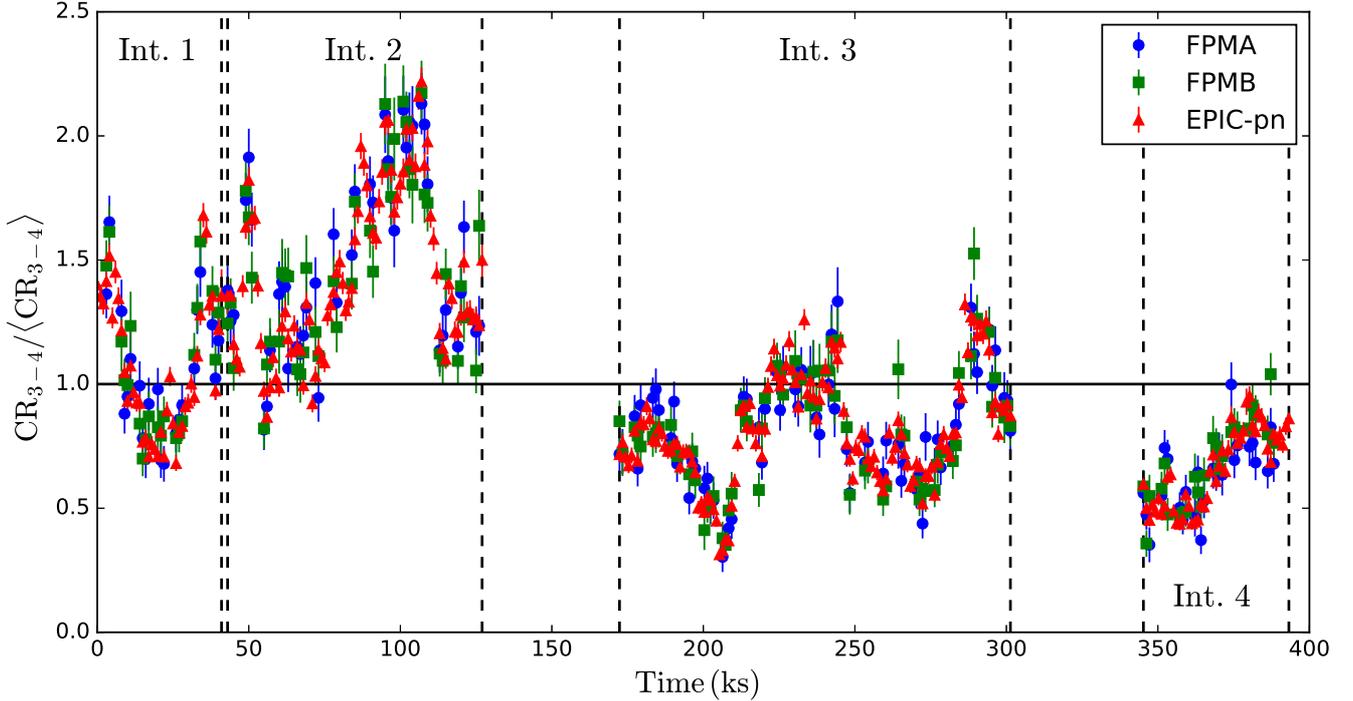}
\caption{The 3--4 keV band EPIC-pn, FPMA, and FPMA (normalized) light curves 
(red triangles, blue circles, and green squares, respectively). The bin size is 1\,ks in all cases and the vertical lines indicate the four intervals when both {\it XMM-Newton} 
and {\it NuSTAR} were observing the source. Time is measured from the start of {\it XMM-Newton} observations.}
\label{fig:lightcurve}
\end{figure*}

Figure\,\ref{fig:lightcurve} shows the {\it XMM-Newton} and {\it NuSTAR} light curves in the 3--4\,keV band (chosen to be the reference band; see next Section), normalized to the mean average count rate. We plot data during the four time periods when both satellites were observing the source (we considered data from these periods only, by merging the good time intervals tables of the two satellites using the {\tt FTOOLS} command {\tt MGTIME}). This figure shows the large variability range of the source (the max-to-min flux ratio is $\sim 7$) but also the consistency between the instruments. 

\section{Flux-flux analysis}
\label{sec:FFA}
\subsection{Choice of the energy bands}
\label{subsec:refband}

The first task in the flux-flux analysis is to define the reference band. Ideally, the flux in this band should be representative of the X-ray primary emission mainly, and should have the largest possible signal-to-noise ratio. In our case this band should also be common in both {\it XMM-Newton} and {\it NuSTAR} data. For these reasons, we chose 3--4\,keV as the reference band. Table\,\ref{table:log} lists the net exposure time and the average 3--4\,keV count rate for each of the 4 time intervals and for the various detectors.

To construct the FFPs at energies above 4 keV (the high-energy FFPs, hereafter) we divided the 4--40\,keV band into 10 sub-bands. The first five were common to both {\it XMM-Newton} and {\it NuSTAR}, with $\Delta E =1$\,keV in the energy range 4--8\,keV, and $\Delta E =2$\,keV for the fifth sub-band (8--10\,keV). Using data from these bands and the reference band we constructed FFPs (plotted in Fig.\,\ref{figapp:commFFPs}). At energies larger than 10\,keV, we used {\it NuSTAR} data only (Fig.\,\ref{figapp:nustarFFPs}). We considered two sub-bands with $\Delta E =2$\,keV. Then we chose a width of $\Delta E=3$\,keV and  5\,keV for the following two sub-bands. We also considered the light curve in the 25--40\,keV sub-band ($\Delta E=15$\,keV). We did not consider the data at energies higher than 40 keV, because of the rapid decrease of the signal-to-noise ratio at these energies. 

At energies below 3 keV, we extracted {\it XMM-Newton} 
light curves from 7 sub-bands in the energy range 
0.3--1\,keV with a width of $\Delta E =0.1$\,keV. Then 
we considered two sub-bands with $\Delta E = 0.3$\,keV, 
one with $\Delta E = 0.4$\,keV, and $\Delta E=1$\,keV for the last 
sub-band (2--3 keV). Using these light curves, and the reference band, we constructed  
the low-energy FFPs (plotted in Fig.\,\ref{figapp:lowEFFPs}). 

\begin{table}
\centering
\caption{Net exposure time and the average count rate in the 
3-4\,keV band for the various time intervals and 
instruments considered in this work.}
\begin{tabular}{cccc}
\hline \hline
Int.	&	Exp. time (ks)	&	 \multicolumn{2}{c}{$ {\rm \langle CR_{3-4} \rangle\,(Count\,s^{-1}) }$ }	\\\cline{3-4}\\[-0.2cm]
	&	EPIC-pn/FPMA,B	&		EPIC-pn		&		FPMA(B)		\\  \hline 	
1	&	41/37	&	1.23	$ \pm $	0.05	&	0.21(0.22)	$ \pm $	0.01		\\[0.2cm]   	
2	&	84/83	&	1.66	$ \pm $	0.04	&	0.30(0.30)	$ \pm $	0.01		\\[0.2cm]   	
3	&	129/129	&	0.93	$ \pm $	0.02	&	0.17(0.17)	$ \pm $	0.01		\\[0.2cm]   	
4	&	48/43	&	0.76	$ \pm $	0.03	&	0.13(0.14)	$ \pm $	0.01		\\[0.2cm]   \hline \hline	
\end{tabular}
\label{table:log}
\end{table}

\subsection{Choice of the time bin size}
\label{sec:timebin}
 
The time bin size of the light curves, $\Delta t_{\rm bin}$, 
plays a significant role in the FFP analysis \citep{Kam15}. 
To investigate this issue, we used {\it XMM-Newton} and {\it Nustar} light curves with 
$\Delta t_{\rm bin} = 100$\,s, 1\,ks, and 5.8\,ks (equal to 
the {\it NuSTAR} orbit)  
to create the low and high-energy 
FFPs (the 100s, 1ks, and 5.8ks FFPs, hereafter). We fitted 
them with a power-law plus constant (PLc) model of the form,
\begin{equation}
y = A_{\rm PLc}x^{\beta} + C_{\rm PLc},
\label{eq:PLc}
\end{equation}
\noindent
($x$ in this, and all equations hereafter, represents the count rate in the reference band). We used the  {\tt MPFIT}\footnote{\url{http://code.google.com/p/astrolibpy/source/browse/trunk/}}  package \citep{Mark09}, taking into account the errors on the $y$-axis only. 

In general, the best-fit parameters in the case of the 1 and 5.8\,ks high-energy FFPs are consistent with each other. This is not the case with the low-energy FFPs. This is similar to what was observed in IRAS~13224--3809 \citep{Kam15} and suggests that the intrinsic FFPs are not linear at energies below $\sim 2-3$ keV (see \S\,\ref{subsec:lowEFFP}).  The model parameters from the best-fits to the 100\,s binned FFPs are significantly different, at all energies. As we demonstrate in Appendix\,\ref{app:poisson}, this discrepancy is due to Poisson noise effects, which become significant when the count rate is low and $\Delta t_{\rm bin}$ is small. We find that the average number of counts per bin in each light curve should be larger than $\sim 200$ photons in order to be able to determine the intrinsic FFP shape, without any distortions due to Poisson noise. Given this result, and the disagreement between the 1\,ks and 5.8\,ks results in the low-energy FFPs, we decided to study the FFPs which are constructed with the use of the 1\,ks binned light curves at all energy bands, except the two highest {\it NuSTAR} energy bands, where we used the 5.8\,ks binned light curves (to satisfy the high count rate criterion). 

\subsection{The high-energy flux-flux plots}
\label{subsec:highE-FFP}

We fitted the high-energy FFPs with the PLc model (eq.\,\ref{eq:PLc}). We fitted both the data of the four time intervals shown in Fig.\,\ref{fig:lightcurve} separately, and the data from all intervals combined together. The fits were statistically accepted in all cases, and the best-fit slopes were consistent with one (at all energies). This result suggests that a straight line can also fit the FFPs. So we re-fitted them with a linear model of the form, 
\begin{equation}
 y = A_{\rm L} x + C_{\rm L},
 \label{eq:linear}
\end{equation}
\noindent
using the 
{\tt MPFFITEXY} routine \citep{Williams10} which 
takes into account the errors on both $x$ and $y$ variables. Tables \ref{table:XMM-highE} and \ref{table:Nustar-highE} in Appendix\,\ref{app:tables} list the best-fit results to the individual and the combined FFPs. The solid lines in Fig.\,\ref{figapp:commFFPs} and \ref{figapp:nustarFFPs} show the best-fit lines to the combined high-energy FFPs. 

\begin{figure}
\includegraphics[width = 0.5\textwidth]{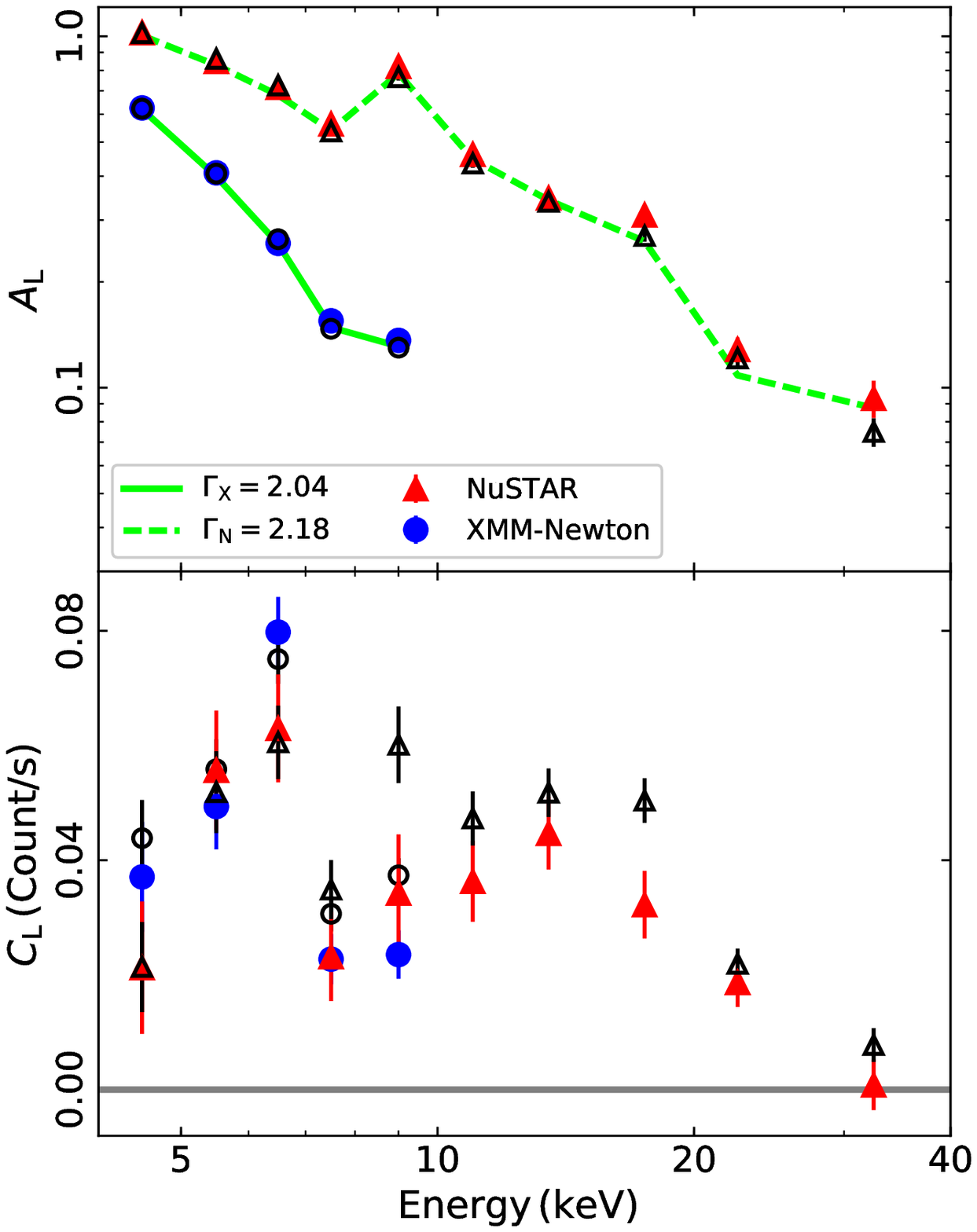}
\caption{The best-fit $A_{\rm L}$ 
and $C_{\rm L}$ (upper/lower panel) parameters derived by fitting a 
linear model to the {\it XMM-Newton} (circles) 
and {\it NuSTAR} (triangles) high-energy FFPs. Filled/empty symbols show 
$A_{\rm L,wm}, C_{\rm L,wm}$ and $A_{\rm L,all},C_{\rm L,all}$, respectively. The solid/dashed lines in the top panel indicate the {\it XMM-Newton} and {\it Nustar} $A_{\rm L}$ model values, assuming a 
power-law spectrum that varies in normalization only (see Section\,\ref{subsec:highE-FFP}).}
\label{fig:weighted-highE}
\end{figure}

The resulting $A_{\rm L}$ and $C_{\rm L}$ values from the best-fits to the FFPs of the individual intervals were consistent within the errors, at all energy bands. Filled symbols in Fig.\,\ref{fig:weighted-highE} show their weighted mean ($A_{\rm L,wm}$ and $C_{\rm L,wm}$) plotted as a function of the mean energy of each energy bin. Empty symbols in the same figure show the best-fit $A_{\rm L,all}$ and $C_{\rm L,all}$ values we get when we fit the combined FFPs (using the data from all four segments). They are consistent with $A_{\rm L,wm}$ and $C_{\rm L,wm}$ (within $3\sigma$). Since the errors of $A_{\rm L,all}$ and $C_{\rm L,all}$ are smaller than the errors of $A_{\rm L,wm}$ and $C_{\rm L,wm}$, we will use the former in our analysis. 

\begin{figure}
\centering
\includegraphics[scale = 0.48]{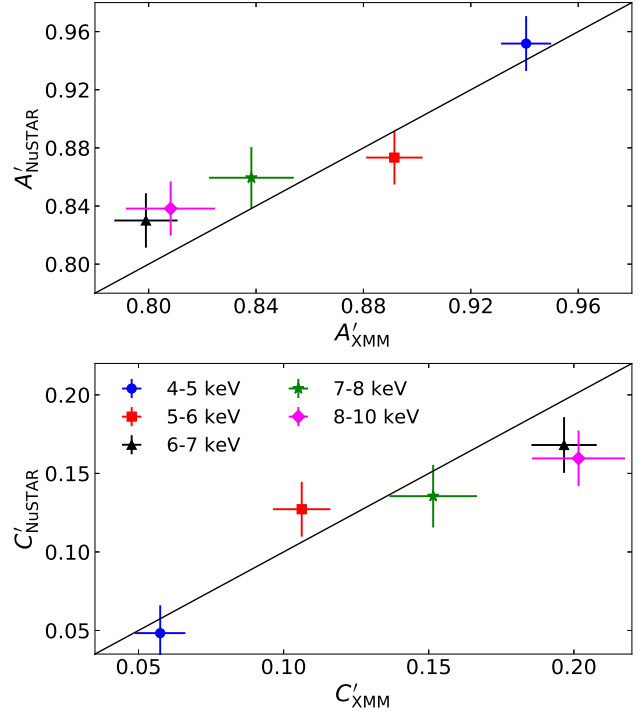}
\caption{Plot of the {\it NuSTAR} versus the {\it XMM-Newton} normalized $A_{\rm L,all}$ (top panel) and $C_{\rm L,all}$ (bottom panel) best-fit  values (see \S\,\ref{subsec:highE-FFP}).
The straight, solid lines indicate the one-to-one relation.}
\label{fig:XNcomp}
\end{figure}

In order to show the consistency of the results derived 
from the {\it XMM-Newton} and {\it NuSTAR} FFPs, we can re-write 
eq.\,\ref{eq:linear} as follows, 
\begin{equation}
\frac{y}{\langle y \rangle} = \frac{A_{\rm L} \langle x \rangle }{\langle y \rangle} \frac{x}{\langle x \rangle} + \frac{C_{\rm L}}{\langle y \rangle},
\end{equation}
where $\langle y \rangle$ and $\langle x \rangle$ are 
the mean count rates. Figure\,\ref{fig:XNcomp} shows 
the normalized {\it NuSTAR} best-fit values (i.e. $A'=A_{\rm L,all}\langle x \rangle/\langle y \rangle$ and $C'=C_{\rm L,all}/\langle y \rangle$),  
versus the respective {\it XMM-Newton} values. 
This plot shows that the results from the analysis 
of the {\it XMM-Newton} FFPs are consistent with those from the 
{\it NuSTAR} FFPs.

The best-fit model constants, $C_{\rm L}$, are significantly larger than zero, even at the highest energy band. This result suggests the presence of a spectral component which is not variable, at least on time scales comparable to the duration of the MCG-06-30-15 observations ($\sim 4.5$ days). Secondly, the high-energy FFPs are well described by a straight line. This is consistent with the hypothesis of a power-law like X--ray continuum which varies in normalization only. In this case, the slope of the line which fits the FFPs,  $A_{\rm L}$, should be equal to the ratio of $y$ over $x$. 

To investigate this issue further, we created fake power-law spectra using the XSPEC \citep{Arn96} command {\tt FAKEIT}, assuming an absorbed PL model with $\Gamma$ in the range 
1.95--2.2, with a step of $\Delta\Gamma=0.01$. We considered only Galactic absorption in the line of sight of the source \citep[$N_{\rm H}=3.92 \times 10^{20}\,{\rm cm^{-2}}$;][]{Kal05}, 
and the response matrices of EPIC-pn and FPMA/B. We estimated the expected count rate 
in each one of the high energy sub-bands, and we computed their ratio over the 3--4 keV model count rate. In this way, we were able to compute $A_{\rm L,mod}$, and then ``$A_{\rm L,mod}-$vs--Energy" data sets for each $\Gamma$ value. 

Then we fitted the observed $A_{\rm L,all}-E$ data (empty symbols in Fig.\,\ref{fig:weighted-highE}) to the $A_{\rm L,mod}-E$ lines. We found that the observed $A_{\rm L}$'s are best reproduced in the case when  $\Gamma_{\rm X} =2.04\pm 0.02$ ($\chi^2_{\rm X}/{\rm degrees\, of\, freedoom\, (dof)}=7.4/4$), and $\Gamma_{\rm N} = 2.18\pm 0.02$ ($\chi^2_{\rm N}/{\rm dof}=22/9$) for the {\it XMM-Newton} and {\it NuSTAR} FFPs\footnote{The difference between the best-fit  $\Gamma_{\rm X}$ and $\Gamma_{\rm N}$ slopes ($\Delta \Gamma = 0.14 \pm 0.03$) should be representative of the inter-calibration uncertainties between EPIC-pn and FPMA/B. For example, the difference we observe is consistent with the $\Delta\Gamma$ differences between the two instruments that  \cite{Madsen2015} reported.}, respectively. The {\it XMM-Newton} and {\it NuSTAR} best-fit $A_{\rm L,mod}-E$ models are plotted with the solid and dashed lines, respectively, in Fig.\,\ref{fig:weighted-highE}. . We note that the best-fit $A_{\rm L,mod}-E$ lines do not give a statistically accepted fit to the data ($\chi^2_{\rm X+N}=29.4/13$ dof, $p_{null}=5.7\times 10^{-3}$). The weighted mean of the residuals ratio ($|(A_{\rm L,mod}-A_{\rm L,all})/A_{\rm L,all}|$) over the 4--40 keV band is $(1.96 \pm 0.49)\%$. Therefore, a PL component which varies in normalization accounts for most, but not all, of the observed variations. We further discuss this issue in \S\,\ref{subsec:Spec-highE}.

\subsection{The low-energy flux-flux plots}
\label{subsec:lowEFFP}

As with the high-energy FFPs, first we fit a PLc model to the low-energy FFPs of the 
individual time intervals. Figure\,\ref{figapp:lowEFFPs} in Appendix\,\ref{app:plots} shows 
the resulting best-fit PLc models. The best-fit results, the mean 
value of the best-fit parameters, and the best-fit parameters obtained 
by fitting all the data together are listed in 
Table\,\ref{table:XMM-lowE}. The model parameters from the best-fits to the individual time intervals 
were consistent with each other, in all bands. However, contrary to the  
high-energy FFPs, the best-fit values derived 
by fitting all the data together do not agree with 
the mean value of parameters obtained by fitting 
the FFPs of the individual time intervals. 

\begin{figure}
\centering
\includegraphics[width = 0.45\textwidth]{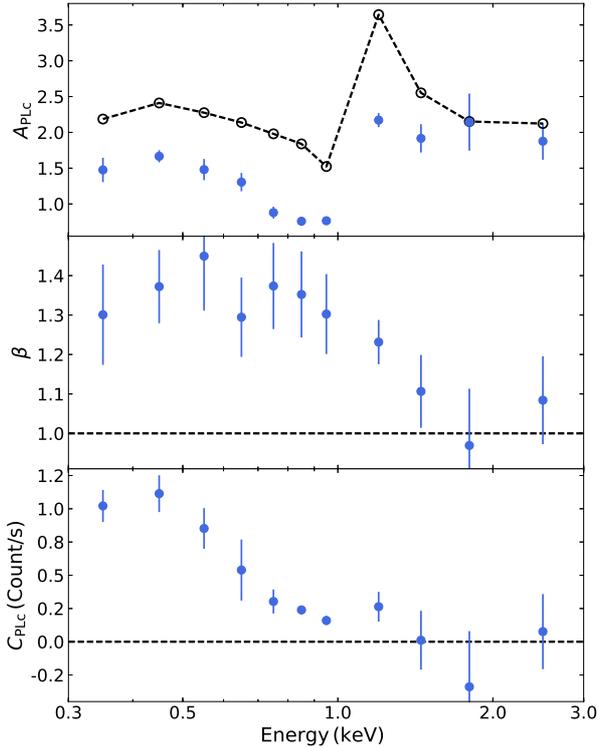}
\caption{Mean best-fit $A_{\rm PLc}$ (upper panel), 
$\beta$ (middle panel) and $C_{\rm PLc}$ (bottom panel) 
values. Empty circles, in the upper panel, show the 
predicted $A_{PLc}$ assuming a PL spectrum with 
$\Gamma_{\rm X} = 2.04$ and a variable normalization.  
(see Sec.\,\ref{subsec:highE-FFP} for details).}
\label{fig:lowE-mean}
\end{figure}

Strictly speaking, the PLc model is not statistically accepted, neither when we fit the individual nor the combined low-energy FFPs. The residual plots show significant, random data fluctuations around the best-fit models, indicative of short-amplitude, fast variations in the low energy bands which are independent of the continuum variations. When we fit a straight line to the best-fit residuals of the individual  FFPs, the best-fit slope turns out to be consistent with zero. This suggests that the PLc model represents rather well the general trend in the low-energy FFPs. It takes account of most of the observed variations in the soft bands, and does not result in any large-scale, systematic trends in the residual plots. On the other hand, the residuals from the best-fits to the combined FFPs show systematic trends. We therefore accept the best-fit results to the individual FFPs as representative of the low energy FFPs. Since the best-fit parameters are consistent (within $3\sigma$) at all low-energy FFPs, we use  
their arithmetic mean\footnote{Due to the large $\chi^2$ values, the error of the best-fit parameters 
does not represent their real uncertainty. For that 
reason we considered the arithmetic mean values of  
the best-fit parameters for the low-energy FFPs.} in our analysis. Filled symbols in Fig.\,\ref{fig:lowE-mean} show the mean model parameters plotted as a function of the centroid energy of each energy bin.

The best-fit model slopes (middle panel in Fig.\,\ref{fig:lowE-mean}) are significantly larger than one at energies below $\sim 1.6$\,keV. Non-linear FFPs can be produced by intrinsic spectral slope variations, as demonstrated by \cite{Kam15}. However, these authors showed that $\Gamma$ variations result in FFP slopes which are flatter than one. In addition, the high-energy FFPs argues against intrinsic $\Gamma$ variations. We therefore conclude that the non-linear FFPs are not the result of spectral slope variations.

The magenta solid lines in Fig.\,\ref{figapp:lowEFFPs} show the expected FFPs assuming a power-law spectrum with $\Gamma_{\rm X} = 2.04$, which varies only in normalization, as is the case with the high-energy FFPs (the predicted FFP lines are plotted assuming the Galactic absorption, only). The open circles in the top panel of Fig.\,\ref{fig:lowE-mean} show the resulting $A_{\rm PLc}$. At energies $\sim 1.6-3$ keV, the observed FFP slopes are consistent with one, and the observed $A_{\rm PLc}$ are consistent with the predicted values. Not surprisingly, the magenta solid lines are also (broadly) consistent with the observed FFPs. We therefore conclude that the FFPs down to $\sim 1.6$ keV are consistent with a power-law spectrum with $\Gamma_{\rm X} \sim 2$, which varies only in normalization. 

The observed FFPs are {\it below} the magenta solid lines at energies between $\sim 0.6-1.6$\,keV.    
Furthermore, the observed $A_{\rm PLc}$ are below the expected values at all energies below $\sim 1.6$ keV. This result suggests that the count rate in these energies is smaller than what we would expect based on the variable PL model that is consistent with the high-energy FFPs (even when we take into account the Galactic absorption).  The lower than expected count rate  can be explained by the well-known variable warm absorber in MCG-6-30-15, which affects mainly the low energy spectrum of the source. At the same time, if the absorber is variable, it can result in FFP slopes which are steeper than one (as we show in Appendix\,\ref{app:warmabs}). 

The best-fit model constants ($C_{\rm PLc}$) are positive at all energies below $\sim 1$\,keV (bottom panel in Fig.\,\ref{fig:lowE-mean}). This is indicative of the presence of a spectral component at low energies which does not vary on time scales shorter than the duration of the observations. This agrees with the fact that, despite the warm absorption, the observed FFPs are above the predicted ones (magenta line) in the 0.3--0.6 keV range. This can only be explained by the presence of an extra spectral component (in addition to the variable PL and the warm absorber).

\section{Discussion}
\label{sec:disc}

\subsection{Absorption induced X--ray continuum variability}

The fact that a straight line fits well the high energy FFPs provides a model independent evidence against variable, clumpy absorption dominating the X--ray variability in MCG-6-30-15. If that were the case, the observed count rate, $y(t)$, at energy $E_y$, would be equal to:
\begin{equation}
y(t) = \left\{\prod_{i=1}^{N} \exp\left[-n_{\rm H,i}(t)\sigma(E_y)\right]\right\} AE_y^{-\Gamma},
\end{equation} 
\noindent
assuming $N$ obscuring clouds, each one with equivalent hydrogen column, $n_{H,i}(t)$, which is variable in time, while the X--ray continuum spectrum remains constant ($\sigma(E_y)$ is the photo-electric cross-section). The above equation becomes, 
\begin{equation}
y(t) =\exp\left\{\left[-\sum_{i=1}^{N}n_{\rm H,i}(t)\right]\sigma(E_y)\right\} AE_y^{-\Gamma},
\label{ynh}
\end{equation}
\noindent
and should also hold for the count rate at energy $E_x$,
\begin{equation}
x(t) = \exp\left\{\left[-\sum_{i=1}^{N} n_{\rm H,i}(t)\right]\sigma(E_x)\right\} AE_x^{-\Gamma}.
\label{xnh}
\end{equation}
\noindent
We can solve for $\left[-\sum n_{\rm H,i}(t)\right]$ using eq.\,\ref{xnh}, and substitute it in eq.\,\ref{ynh} in order to reach the following relation between the count rates in the two bands,
\begin{equation}
y = Cx^{\beta},
\label{xynh}
\end{equation}
\noindent
wehere $C$ is a constant, and $\beta=\sigma(E_y)/\sigma(E_x)$. Equation \ref{xynh} predicts a  non linear relation between $y$ and $x$, contrary to our results. Even if $N$ varies with time, eq.\,\ref{xynh} should still hold. Therefore, our results show that the hypothesis that the X--ray variability in MCG-6-30-15 is due to variable absorption only (on the time scales we probe, at least) is not valid. 

\subsection{The constant high energy X--ray component}
\label{subsec:highE-constant}

\begin{figure}
\centering
\includegraphics[width = 0.49\textwidth]{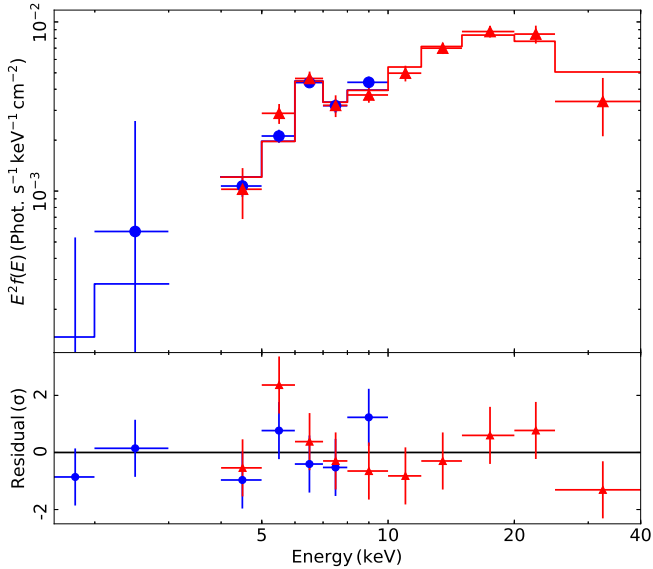}
\caption{The energy spectrum of 
the constant component (top panel) fitted with a 
neutral reflection model together with the corresponding 
residuals (bottom panel) for {\it XMM-Newton} (blue circles) 
and {\it NuSTAR} (red triangles).}
\label{fig:pexmonSpec}
\end{figure}

\begin{table}
\centering
\caption{The best-fit parameters obtained by fitting the high energy constant component with {\tt pexmon}.}
\begin{threeparttable}
\begin{tabular}{lcccc}
\hline \hline
$\Gamma$	&	$	2.06	_{-	0.19	}^{+	0.17	}$	&	$	2.03	_{-	0.19	}^{+	0.17	}$	&	$	1.99	_{-	0.20	}^{+	0.18	}$	&	$	1.91	_{-	0.20	}^{+	0.19	}$ \\[0.1 cm] 
$E_{\rm cut}$ &	$	27	_{-	7	}^{+	12	}$	&	$	26	_{-	7	}^{+	12	}$	&	$	26	_{-	7	}^{+	12	}$	&	$	25	_{-	6	}^{+	11	}$\\
(keV) & & & & \\[0.1 cm]
$A_{\rm Fe}$&	$	0.26	_{-	0.04	}^{+	0.05	}$	&	$	0.26	_{-	0.04	}^{+	0.05	}$	&	$	0.26	_{-	0.04	}^{+	0.05	}$	&	$	0.27	_{-	0.04	}^{+	0.05	}$\\[0.1 cm]
$i(^\circ)$	&	$	{0^f					}$	&	$	{30^f					}$	&	$	{45^f					}$	&	$	{60^f					}$ \\[0.1 cm]  
Norm &	$	0.026	_{-	0.007	}^{+	0.009	}$	&	$	0.025	_{-	0.007	}^{+	0.009	}$	&	$	0.025	_{-	0.007	}^{+	0.009	}$	&	$	0.026	_{-	0.007	}^{+	0.009	}$ \\ \hline 
$\chi^2 / {\rm d.o.f.}$	&		14.25/12						&		14.19/12						&		14.43/12						&		14.8/12	\\[0.1cm] \hline\hline
\end{tabular}
\begin{tablenotes}
    \item[$f$] Fixed.
\end{tablenotes}
\label{table:pexmon}
\end{threeparttable}
\end{table}

The linear model defined by eq.\,\ref{eq:linear} consists of two terms. The $C_{\rm L}$ term should be representative of a spectral component, which is not variable (at least over the sampled time scales). We used the  best-fit $C_{\rm L,all}$ values and the {\tt FTOOLS} command {\tt ascii2pha} to construct the spectrum of this component at energies above 1.6 keV (Fig.\,\ref{fig:pexmonSpec}). We fitted the spectrum with the neutral reflection model {\tt pexmon} \citep{Nan07}. We fixed the reflection 
fraction to one and the abundance of heavy elements to solar but we let the iron abundance and the cutoff energy free to vary. We kept all the parameters tied between the {\it XMM-Newton} and {\it NuSTAR} spectra, but we included a multiplicative cross-calibration constant that we fixed to unity for the {\it XMM-Newton} spectrum and we let it free to vary for the {\it NuSTAR} spectrum. We found it consistent with one , for all the cases that we considered. 

The best-fit results are listed in Table~\ref{table:pexmon} for various inclinations, up to 60 degrees. The model fits well the data in all cases, which implies that we cannot constrain the inclination. The photon index is consistent with 2 (within the errors), and the iron abundance is subsolar in all cases. We also found a low value for the high-energy cutoff, similar to the one found for the variable component in \S\,\ref{subsec:Spec-highE} (the 3\,$\sigma$ upper limit is 120\,keV). Our results indicate that the constant component can result from reflection off neutral material. This component is constant over at least $\sim 4.5$ days, which places a lower limit on the distance of the reflector from the central source. Assuming that the BH mass is $M_{\rm BH} \simeq 1.6\times 10^6\,M_{\rm \odot}$ \citep{Bentz16}, this implies that the reflecting material is located at a distance $D \geq 5\times 10^4\,r_{\rm g}$ ($r_{\rm g}=GM_{\rm BH}/c^2$, is the gravitational radius). This is $\sim 1.7$ times larger than the broad line region radius in this source \citep{Mari14}.

\subsection{The constant low energy X--ray component}
\label{sec:spec-lowE}

\begin{figure}
\centering
\includegraphics[width = 0.49\textwidth]{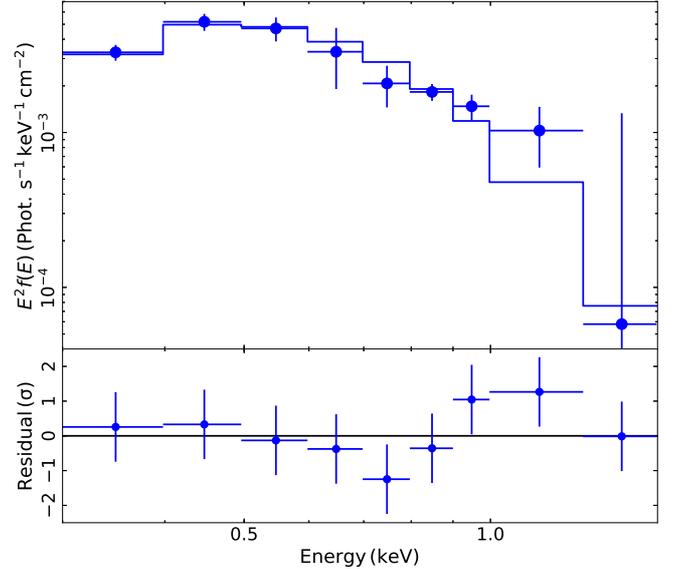}
\caption{The energy spectrum of 
the low-energies constant component fitted with a 
black-body model (solid line, top panel), and the 
corresponding best-fit residuals (bottom panel).}
\label{fig:BBSpec}
\end{figure}

The model defined by eq.\,\ref{eq:PLc}, which fits well the low-energy FFPs, also consists of two terms. The $C_{\rm PLc}$ term could be representative of a low-energy spectral component which remains constant on time-scales of a few days (at least). However, this is not straight forward in this case. The soft X-ray spectrum of MCG--6-30-15 is charaterized by complex and variable warm absorption. We demonstrate in Appendix\,\ref{app:warmabs} that variable warm absorption can result in non-linear FFPs at low energies, with slopes steeper than one, as observed. The simulated FFPs are well fitted by a PLc model, with either positive or negative constants, $C_{\rm PLc,sim}$. In the cases that we considered, the absolute value of these constants is much smaller than the constants we measure in the observed low-energy FFPs, $C_{\rm PLc,obs}$. Although we cannot prove that this will always be the case, it is possible that  $C_{\rm PLc,obs}$ are indicative of a spectral component which does not vary, at least over the duration of the observations. 

We used the best-fit $C_{\rm PLc,obs}$ values listed in Table\,\ref{table:XMM-lowE} (and the  {\tt FTOOLS} command {\tt ascii2pha})  to construct the low-energy, constant spectral component of MCG-6-3015 (plotted  in Fig.\,\ref{fig:BBSpec}). We fit the spectrum with an absorbed blackbody (BB) spectrum, taking into account the Galactic absorption only. The fit (blue solid line in Fig.\,\ref{fig:BBSpec}) is statistically accepted ($\chi^2/{\rm d.o.f.}=4.7/7$). The best-fit temperature and normalization are $kT_{\rm BB} = 100 \pm 6$\,eV and $N_{\rm BB} = (1.99 \pm 0.3)\times 10^{-4}$, respectively.

Such a component could be due to the intrinsic emission of the inner disc. In this case, this component should be variable on the local viscous time scale, which even for a source with a BH mass of the order of a million solar masses could be of the order of many days.  
{In order to investigate the possibility of this component being representative of disc emission, we considered the {\tt optxagnf} model \citep{Done12} which gives the spectral energy distribution of an accretion disc around a rotating SMBH, assuming Novikov-Thorne emissivity. We fitted this model to the data, assuming a BH mass of $1.6\times 10^6\,M_{\odot}$, a spin parameter of 0.998, and the emission from the inner part of the disc only (i.e. we fixed the model parameter $r_{out}$ to 2\,$r_{\rm g}$). The model fits the data well ($\chi^2= 12.9/8$ dof, $p_{null} = 0.11$), with the best-fit Eddington ratio being $\log \lambda_{\rm Edd} = -1.19 \pm 0.02$. We therefore conclude that, the constant component in the soft-band of MCG--6-30-15 can be indicative of the inner disc emission, if the BH is maximally rotating, and the accretion rate is $\sim 6$ per cent of the Eddington limit.}

We note that the best-fit residuals plot in Fig.\,\ref{fig:BBSpec} indicate an absorption feature at energies $\sim 0.6-0.8$ keV. It is not significant but this feature is reminiscent of warm absorption. It suggests that the constant soft spectral component is emitted by a region close to the central source, in agreement with the assumption that this is the intrinsic emission from the inner disc. 

\subsection{The variable X--ray spectral component}
\label{subsec:Spec-highE}

{The $A_{\rm PLc}x^{\beta}$ and $A_{\rm L}x$ terms in eqs.\,\ref{eq:PLc} and \ref{eq:linear} should account for the variable, X--ray continuum spectral component in MCG--6-30-15, at low and high energies, respectively. We considered the mean 3--4\,keV count rate with the 
best-fit $A_{\rm L,all}$ and $A_{\rm PLc}$ values (at energies above and below 1.6 keV, respectively),  and we used the {\tt FTOOLS} command {\tt ascii2pha} to create the spectrum of this component. In principle, we could use any 3--4\,keV count rate value to create the spectrum. We chose the mean so that
the resulting spectrum is representative of the variable component in the average-flux state of the source (during these observations). The high energy variable component, $y_{\rm var,h}$, is plotted with the filled symbols in the top panel of Fig.\,\ref{fig:meanvarspec} (circles and triangles indicate the data using the best-fit {\it XMM-Newton} and {\it NuSTAR}  $A_{\rm L,all}$ values, respectively). The low-energy  variable component,  $y_{\rm var,l}$, is plotted with the open circles in the same panel. 
  
We fitted $ y_{\rm var,h}$ with a PL model, taking into consideration the Galactic absorption in the line of sight of the source. The model provides a rather poor fit to the data  ($\chi^2=30$/15 dof; $p_{null}=0.01$), in agreement with the results we presented in \S\ref{subsec:highE-FFP}.  The weighted mean of the residuals ratio in the 2--10\,keV band is $1.5\pm 0.5\%$. This in agreement with the results from the principle component analysis (PCA) method which reveals that the variability in the normalization of the PL component can account for $\sim 97\%$ of the variability in this source \citep{Parker14, Parker15}.

\begin{figure}
\centering
\includegraphics[width = 0.49\textwidth]{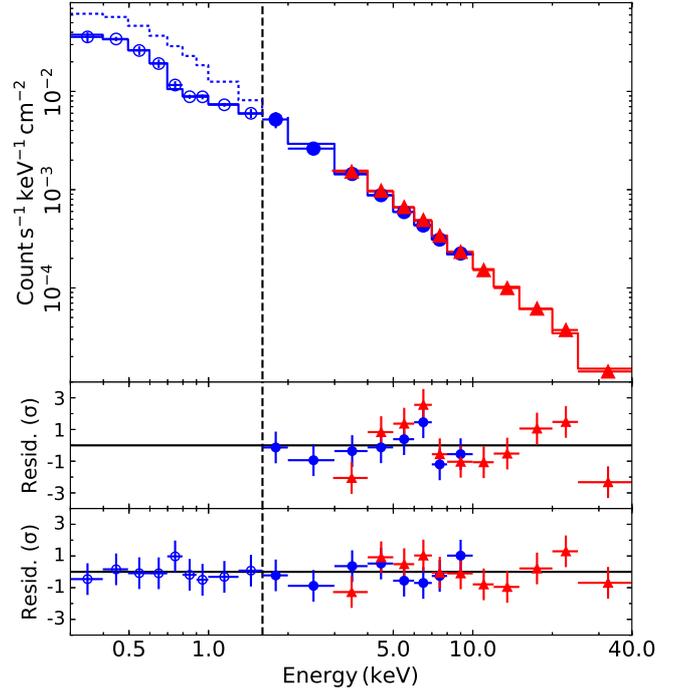}
\caption{{\it Top panel:} The average, X--ray variable spectrum in MCG-6-30-15  using the best model fit results to the {\it XMM-Newton}  and {\it NuSTAR} FFPs (circles and triangles, respectively). 
The vertical dashed line at 1.6 keV indicates the boundary between the high and low energy parts of the spectrum (assuming that the FFPs at energies 1.6--3 keV are similar to the high-energy FFPs, as we argued in \S\,\ref{subsec:lowEFFP}). The solid line above 1.6\,keV indicates the best-fit {\tt relxill} model to the data. The dotted line indicates the extrapolation of the model to lower energies. The solid line below 1.6\,keV indicates the best-fit {\tt zxipcf $\times$ relxill} model to the data. {\it Middle and bottom panels:} The PL  and the  {\tt zxipcf $\times$ relxill} best-fit residuals, respectively (see \S\,\ref{subsec:Spec-highE} for details).}
\label{fig:meanvarspec}
\end{figure}

The best-fit residuals (shown in the middle panel in Fig.\,\ref{fig:meanvarspec}) indicate a deficit at $\sim 3$ keV and an excess at around $\sim 6.5$ and 20~keV, and are  are suggestive of an X-ray reflection component. We therefore re-fitted $ y_{\rm var,h}$ with {\tt relxill} \citep{Daus13, Gar14} (accounting for Galactic absorption). We assumed a maximally spinning black hole, a power-law emissivity profile with $q=3$, and a reflection fraction of 1. We fixed the inner and outer disc radius to the ISCO and to 400\,$r_{\rm g}$, respectively. The model fits the data well ($\chi^2/{\rm dof}=10.8/11$; the best-fit residuals are plotted in the bottom panel of Fig.\,\ref{fig:meanvarspec}). 

The best-fit results are listed in the second column of Table\,\ref{table:parvar}. The best-fit spectral slopes are consistent with the spectral slopes we found in \S\ref{subsec:highE-FFP}. The best-fit PL cut-off energy is rather low when compared to other AGN \citep[e.g.][]{Marinucci16} but it is not well constrained. The $3\sigma$ confidence range is [34--295~keV]. We note that the respective $E_{\rm cut}$ range from the {\tt pexmon} best-fit to the constant component (for all inclinations) is [12--120~keV]. When combined together, the two results  indicate a cut-off energy between 34--120~keV in MCG-6-30-15. We also note that the best-fit iron abundances from the {\tt relxill} fit to the variable component and from the {\tt pexmon} fit to the constant component are not in agreement. We cannot explain this discrepancy. It could either mean that our modelling is not complete, or it may be indicative of the degree that one (or both) of the models approximate well the respective spectral components.

\begin{table}
\centering
\caption{The best-fit {\tt relxill} and {\tt zxipcf} results from the modelling of the variable component in the 1.6--40 keV and the 0.3--40 keV bands (second and third/forth columns, respectively. See \S\,\ref{subsec:Spec-highE} for details).}
\begin{threeparttable}
\begin{tabular}{lccc}
\hline \hline
\multicolumn{4}{c}{\tt relxill}													\\[0.2cm]   
$\Gamma_{\rm X}$	&	$	2.03 \pm 0.03	$	&	$	2.03^f	$	&	$2.12 _{-0.04}^{+0.10}$	\\[0.2cm]   
$\Gamma_{\rm N}$	&	$	2.16 \pm 0.05	$	&	$	2.16^f	$	&	$2.25_{-0.05}^{+0.10}$	\\[0.2cm]   
$i (^\circ)$	&	$	42_{-10}^{+5}	$	&	$	42^f	$	&	$44_{-9}^{+6}$	\\[0.2cm]   
$\log \xi_{\rm d}$	&	$	1.7_{+0.3}^{+0.2}	$	&	$	1.7^f	$	&	$1.69_{-0.40}^{+0.38}$	\\[0.2cm]   
$A_{\rm Fe}$\,(solar)	&	$	1.48_{-0.60}^{+0.89}	$	&	$	1.48^f	$	&	$0.88_p^{+0.56}	$	\\[0.2cm]   
$E_{\rm cut}$\,(keV)	&	$	60_{-15}^{+23}	$	&	$	60^f	$	&	$81_{-64}^{+200}	$	\\[0.2cm]  \hline  
\multicolumn{4}{c}{\tt zxipcf}													\\[0.2cm]    
$N_{\rm H}\,(10^{21} \rm cm^{-2})$	&	$	-	$	&	$	5.1_{-0.6}^{+1.1}	$	&	$7.1_{-1.1}^{+2.3}	$	\\[0.2cm]   
$\log \xi_{\rm abs}$	&	$	-	$	&	$	0.78 \pm 0.10	$	&	$	0.66_{-0.18}^{+0.10}	$	\\[0.2cm]   
CF	&	$	-	$	&	$	0.96_{-0.08}^p	$	&	$	0.88 \pm 0.07	$	\\[0.2cm]   \hline  
$\chi^2/{\rm dof}$	&	$	10.8/11	$	&	$	17/23	$	&	$	10.8/17	$	\\[0.2cm]   \hline \hline
\end{tabular}
\begin{tablenotes}
	\item[$p$] pegged to its maximum/minimum value.
	\item[$f$] fixed.
\end{tablenotes}
\end{threeparttable}
\label{table:parvar}
\end{table}

The extrapolation of the best-fit {\tt relxill} model to low energies ($<1.6$\,keV) is indicated by the dotted blue line in the top panel of Fig.\,\ref{fig:meanvarspec}. The model exceeds the average variable component in this energy range. This is due to the effects of the warm absorber. Hence, we fitted the full band (0.3--40\,keV) variable component  with the model: {\tt zxipcf $\times$ relxill} (accounting for Galactic absorption). First, we fixed the {\tt relxill} parameters to their best-fit values obtained from fitting $ y_{\rm var,h}$. The fit was statistically acceptable ($\rm \chi^2/dof = 17/23$). The best-fit warm absorber parameters are listed in the third column of Table\,\ref{table:parvar}. The best-fit model and the corresponding residuals are shown in the top and bottom panel of Fig.\,\ref{fig:meanvarspec}, respectively. We re-fitted the full band variable spectrum with the same model but letting the {\tt relxill} parameters free. The fit was also acceptable ($\rm \chi^2/dof=11/17$). The best-fit parameters are reported in the last column of Table\,\ref{table:parvar}. There are differences between the best-fit values listed in the first and third columns of Table\,\ref{table:parvar}, notably in the PL spectral slopes, but they are within 2\,$\sigma$.

Our results imply that the observed variations in MCG-6-3015 are due to a PL continuum which is variable in normalization only, and a variable, X--ray reflection component from the (ionized) inner disc. Various studies in the past have detected short delays between the continuum and the soft band variations in this source \citep[e.g.][]{Emma14, Kara14}. Recently, \cite{Epitropakis16} also detected similar delays between the continuum and the iron line variations in MCG--6-30-15. To measure time lags, both the continuum and the reflection components must be variable. Our results confirm this scenario. 
}
\section{Conclusions}
\label{sec:conclusion}

To correctly estimate flux-flux plots, the mean counts per bin in both light curves must be larger than 200 in order to avoid distortions in the FFP shape due to the Poisson noise bias. As long as this criterion is fulfilled, the bin size of the light curves should be as small as possible, in order to avoid further distortions due to binning, in the case when the intrinsic FFP has a non-linear shape. 

{The FFP analysis can provide model independent information on both the constant and variable spectral components in the X--ray spectra of AGN. The latter possibility has not been explored in detail so far, although it has interesting advantages. For example, the FFP shape (linear or power-law like) can show conclusively, and in a model independent way, whether variable absorption operates or not. The spectrum shown in Fig.\,\ref{fig:meanvarspec} is not a traditional, observed spectrum. It is a representation of the spectral energy distribution of the source at a certain flux level, using the results from the FFP analysis. Its energy resolution is low, but it is free of non-variable spectral components that complicate the subsequent model fitting. We could construct these spectra at various flux levels, and study the spectral evolution of the source in this way. We plan to explore in detail this possibility in the future.} Our conclusions from the study of the MCG--6-30-15 FFPs are summarised below. 
\\

\noindent
{\it A) The non-variable, X--ray spectral components in MCG--6-30-15.} 
\\

A1) We detect spectral component(s) that remain constant at least over the duration of the observations we study (i.e. $\sim 4.5$ days). At energies above $\sim 1.6$ keV the constant spectral component is consistent with reflection from cold, neutral material, located more than $5 \times 10^4\,r_{\rm g}$ away from the central source. Our results are consistent with the results of \cite{Tay03}. At energies below $\sim 1.6$\,keV, the constant component is well fitted by a black-body model with a temperature of $\sim 0.1$\,keV. This component cannot correspond to the soft-excess expected from X--ray reflection from a milddly ionized disc, as this should be variable (since the reflection at high energies is variable). It could be due to intrinsic thermal emission from the inner disc itself, if the disc extends to the ISCO around a maximmaly spinnign BH. 

A2) The 2--10 and 2--40\,keV flux of the high energy, constant component is $5\times 10^{-12}$ and $1.9\times 10^{-11}$ $\rm erg\,s^{\rm -1}cm^{-2}$, respectively, which is 10\% and 20\% of the average X--ray continuum flux. The 0.3--1.6~keV flux of the low energy component is $\sim 17\%$ of the average X--ray continuum flux in the same band. These are not negligible fractions so, in addition to a PL continuum plus a relativistically blurred reflection component, modelling of the X--ray spectrum of the source should also add: a) a constant reflection component from cold material, and b)  a constant, blackbody-like component at low energies.
\\

\noindent
{\it B) The variable, X--ray spectral components in MCG--6-30-15}.
\\

B1) The  FFPs at energies above $\sim 1.6$ keV are well fitted with a straight line. This result proves  that: a) there are no spectral slope variations, and b) the observed variations cannot be caused by variations of the number and/or the covering factor of absorbing clouds. These are straight forward results, which do not depend on any assumptions regarding the model fitting of the source's spectrum.

B2) Both the low and the high energy FFPs are fully consistent with a PL continuum, which varies in normalization, plus a variable (on time scales as short as 1\,ks), X-ray reflection component, from ionized material close to the central BH. The variable reflection component is consistent with the detection of ``soft" time lags in this source, since in order to detect delays between two components, both of them must vary. Part of the observed variations at energies below $\sim 1$\,keV are due to variations of the warm absorber. The presence of the variable warm absorber  is supported by the non-linearity of the FFPs\ at energies below 1.6\,keV are non-linear (like IRAS\,13224--3809). 
\\

\noindent
{\it C) The soft excess in MCG-6-30-15}. 
\\

It consists of both a constant and a variable component. Both could originate from the inner disc, as
long as it extends to the ISCO around a fast rotating
BH: the former could be due to the disc's intrinsic
emission, the latter due to X--ray reprocessing (from the same disc region). Using the best-fit results of the constant and variable components, we estimate that the 0.3--1 keV flux of the constant and the variable component, in excess of the PL, are $6.8\times 10^{-12}$ and $4.3\times 10^{-12}$ ergs s$^{-1}$ cm$^{-2}$, respectively. Therefore, $\sim 60$ and 40 per cent of the soft excess flux is due to these two components. We note that that the variable component flux is based on the modeling of the variable component we reported in \S\,\ref{subsec:Spec-highE}, when the source was in its average-flux state during the 2013 observations. Obviously, the contribution of the variable soft excess component (due to X--ray reprocessing) will be larger/smaller during higher/lower flux states of the source.

\section*{Acknowledgements}

We thank the anonymous referee for useful comments. This work made use of data from the {\it NuSTAR} mission, a project led by the California Institute of Technology, managed by the Jet Propulsion Laboratory, and funded by NASA, {\it XMM-Newton}, an ESA science mission with instruments and contributions directly funded by ESA Member States and NASA. This research has made use of the {\it NuSTAR} Data Analysis Software (NuSTARDAS) jointly developed by the ASI Science  Data Center (ASDC, Italy) and the California Institute of Technology (USA).

\bibliographystyle{mnras}
\bibliography{ek-MCG-ref} 

\appendix

\section{The Poisson noise effects to FFPs}
\label{app:poisson}

We chose the 25--40 vs 3--4 keV {\it NuSTAR} FFP  (bottom panel in Fig.\ref{figapp:nustarFFPs}) to investigate the effects of the Poisson noise on the FFPs, because the mean count rate in these bands is the smallest among all FFPs. 
First we created simulated ({\it NuSTAR}) 3--4 keV band count rates assuming a log-normal distribution with mean and standard deviation equal to the mean and standard deviation of the observed count rates in this band. Using the resulting values we computed 25--40 keV band count rates based on the best-fit linear relation we obtained from fitting the observed, 1\,ks binned FFP. We multiplied the count rates in both bands by a factor equal to 1, 2, 3, 4 and $5\times 10^3$, assuming a Poisson distribution, in order to compute the simulated counts. We divided the resulting counts by the respective factor to get the final, simulated count rate in both bands, and we used them to construct 1, 2, 3, 4 and 5\,ks binned, simulated FFPs. Then, we fitted them with a linear model, exactly as we did with the observed FFPs. 

\begin{figure}
\centering
\includegraphics[width = 0.5\textwidth]{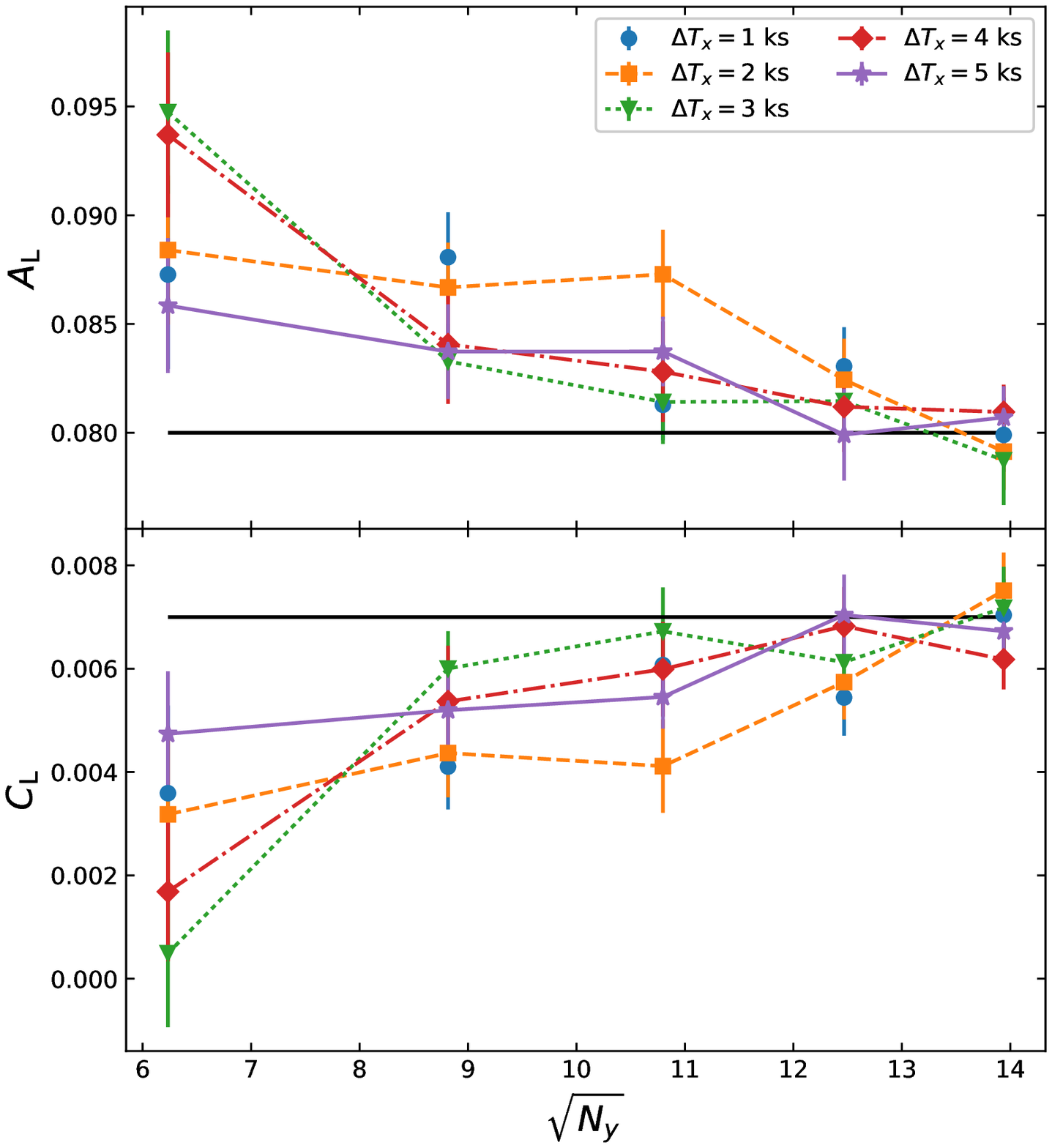}
\caption{The best-fit $A_{\rm L}$ and $C_{\rm L}$ constants (top and bottom panels, respectively) of a linear model fit to simulated FFPs, assuming 1, 2, 3, 4 and 5 ksec binned light curves with a mean count rate equal to that of the 25--40 and 3--4 keV band, {\it Nustar} light curves. The horizontal lines indicate the intrinsic value of the constants.}
\label{fig:T100}
\end{figure}

Figure~\ref{fig:T100} shows the best fit $A_{\rm L}$ and $C_{\rm L}$ values (top and bottom panels, respectively), as a function of the square root of the average counts in the (simulated) 25--40 keV band light curve. The two panels in Fig.\,~\ref{fig:T100} show that we can retrieve the intrinsic $A$ and $C$ values (indicated by the horizontal line in both panels), only when the average counts in the light curve is at least $\sim 200$. The mean count rate in the 3--4 keV {\it NuSTAR} band is ten times larger than the mean count rate in the 25--40 keV band (see Fig.\,\ref{figapp:nustarFFPs}). In fact, the average counts in this band is larger than 200 even if we bin the data into 1\,ks bins. For that reason, the best-fit $A_{\rm L}$ and $C_{\rm L}$ values are consistent (with the error), irrespective of the bin size of the 25--40 keV light curves. However, they approach the intrinsic values only when the average counts in the 25--40\,keV band light curves reaches the limit of 200. 

\cite{Kam15} suggested the use of light curves with the shortest possible bin size in order to recover the intrinsic FFP shape in the case of highly variable sources and an intrinsically non-linear relation. We show here that in doing so, particular case should be given to Poisson noise effects, which can affect the observed shape of the FFPs. Although it is usually assumed that the Poisson distribution approaches the Gaussian distribution when the mean (ie the average counts) is $\sim 20-50$, our results indicate that this assumption is not enough to guarantee the correct estimation of the model parameters when fitting FFPs. We suspect that the reason is due to the nature of the intrinsic count rate distribution. As indicated by the plots in Appendix\,\ref{app:plots}, there is usually a small number of high flux points, which can span a large range in fluxes. It appears that a truly large number of counts per bin is necessary to guarantee a good approximation to a Gaussian (which is symmetric), so as to not bias the best-fit results to the FFPs to steeper (than intrinsic) slopes (and hence smaller constants). 

{We considered light curves affected by Poisson noise, because this is usually the case with X--ray light curves, such as the {\it NuSTAR} light curves. Our conclusions should be largely unaffected by the nature of the experimental noise (be it Poissonian or not): as long as the (mean) signal to noise ratio of the observed light curves, defined for example as the ratio of the mean over the mean error, is larger than $(N=200)/\sqrt{N=200}\sim14$, then the resulting FFPs should not be affected by the effects of the observational noise bias.}

\section{Plots}
\label{app:plots}

\begin{figure}
\centering
{\includegraphics[width=0.225\textwidth]{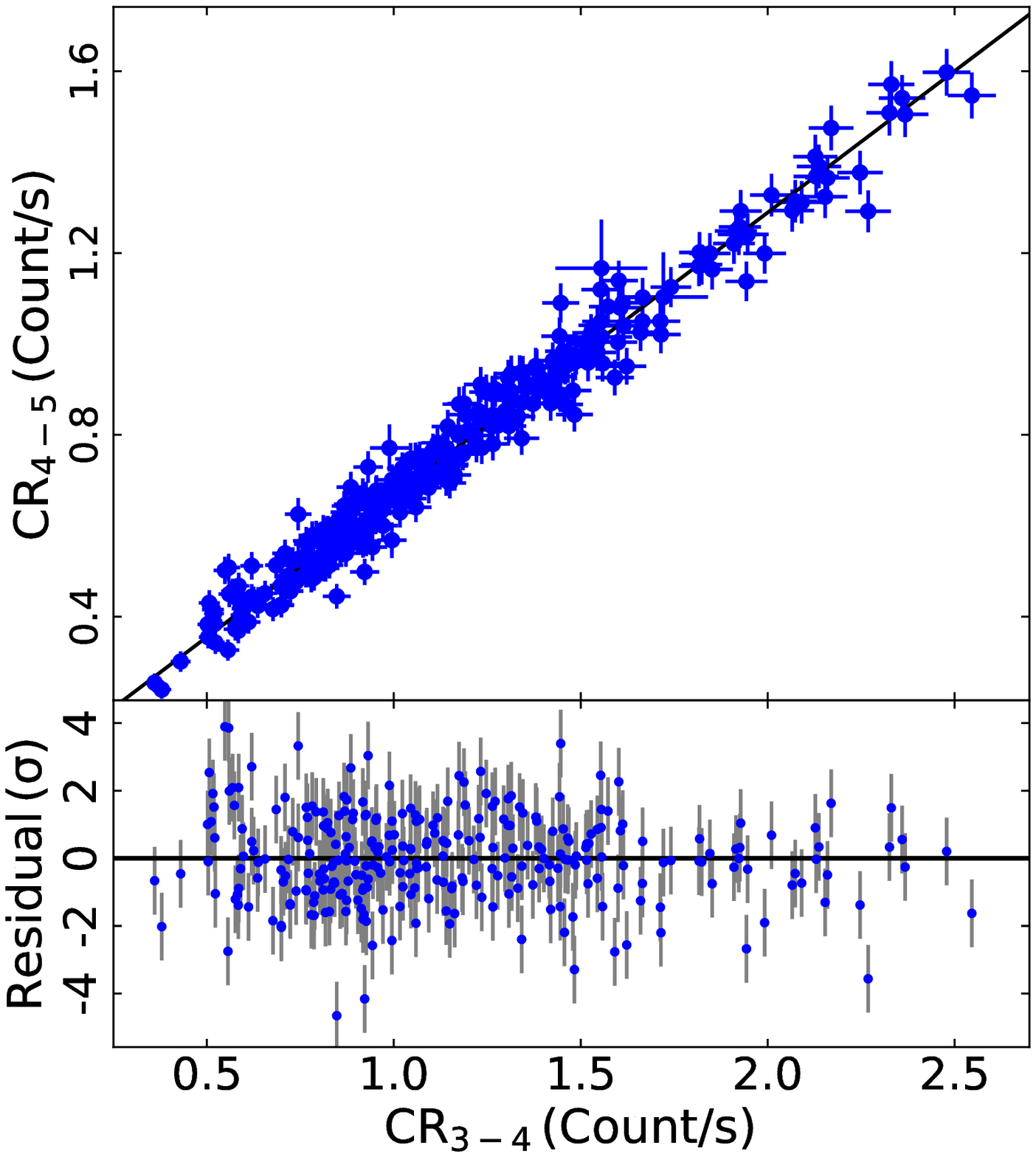}}
{\includegraphics[width=0.225\textwidth]{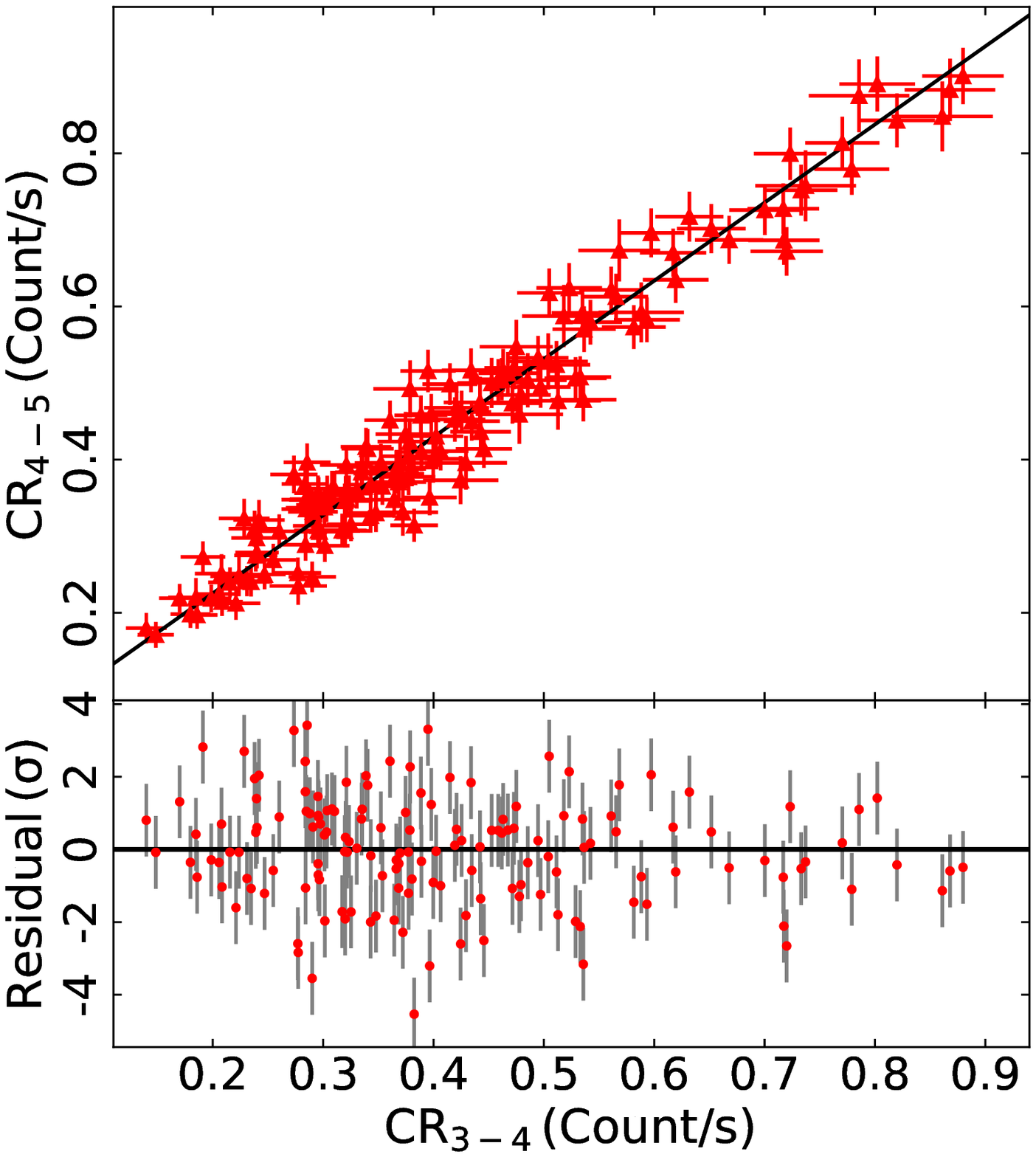}}
{\includegraphics[width=0.225\textwidth]{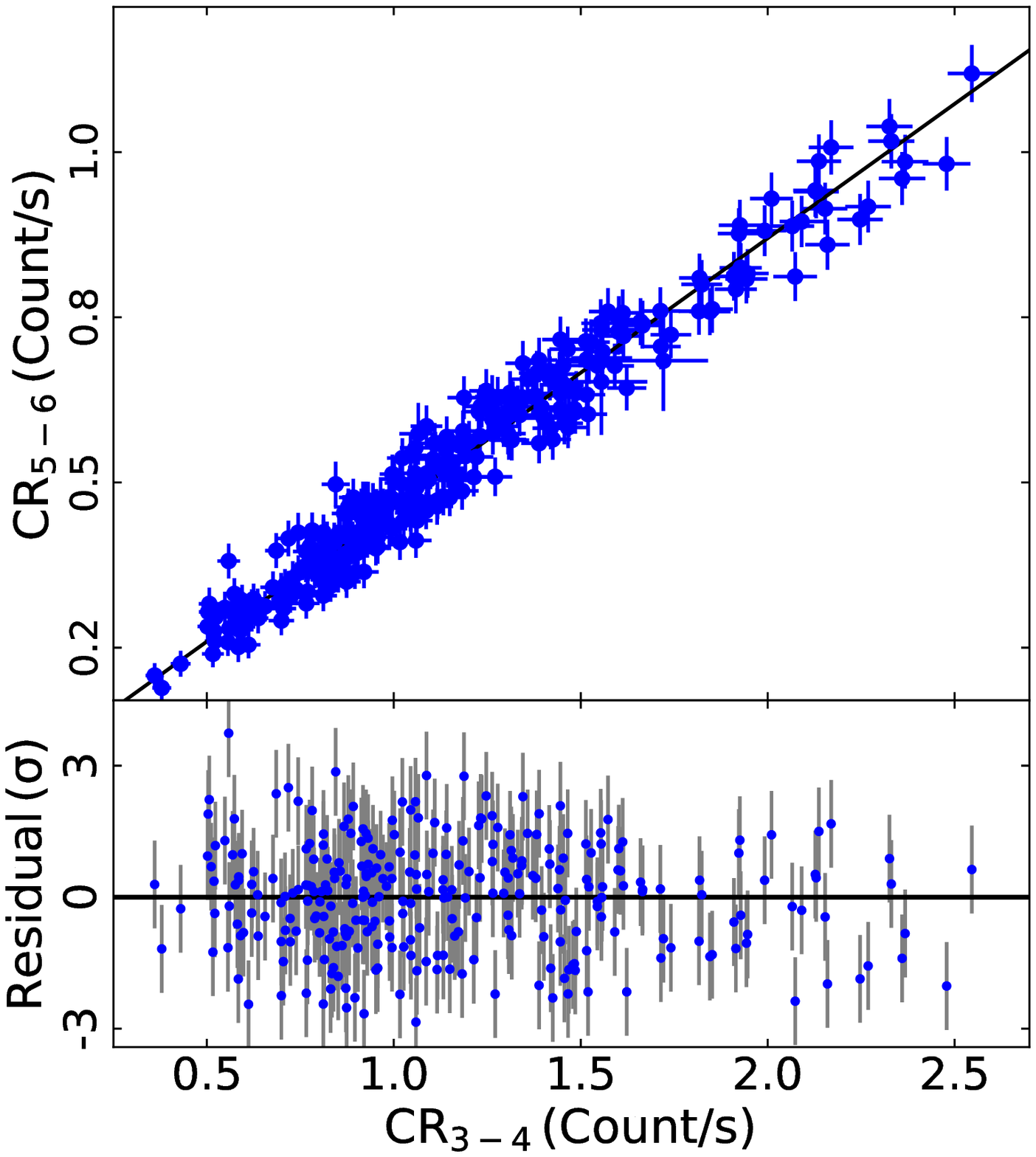}}
{\includegraphics[width=0.225\textwidth]{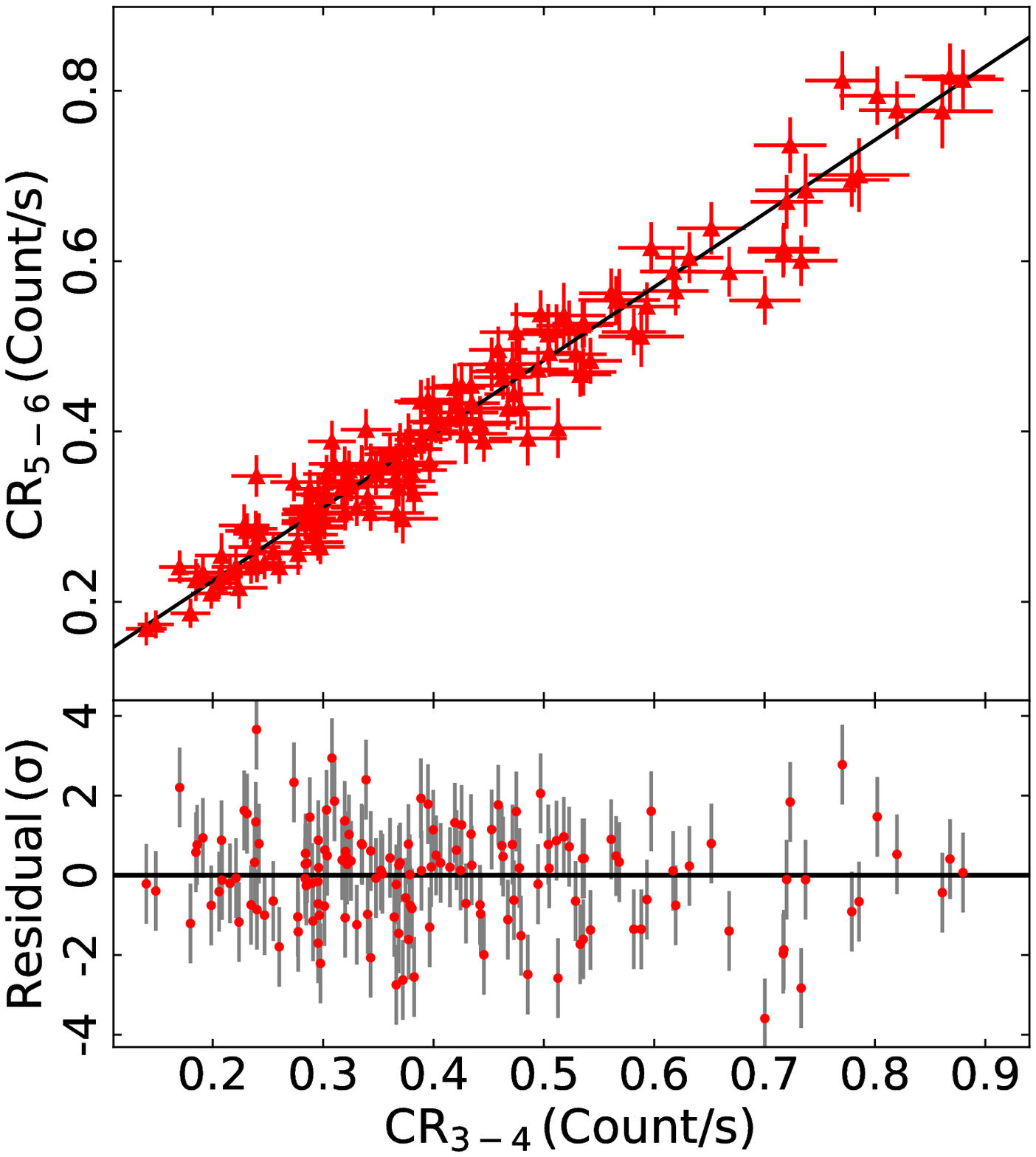}}
{\includegraphics[width=0.225\textwidth]{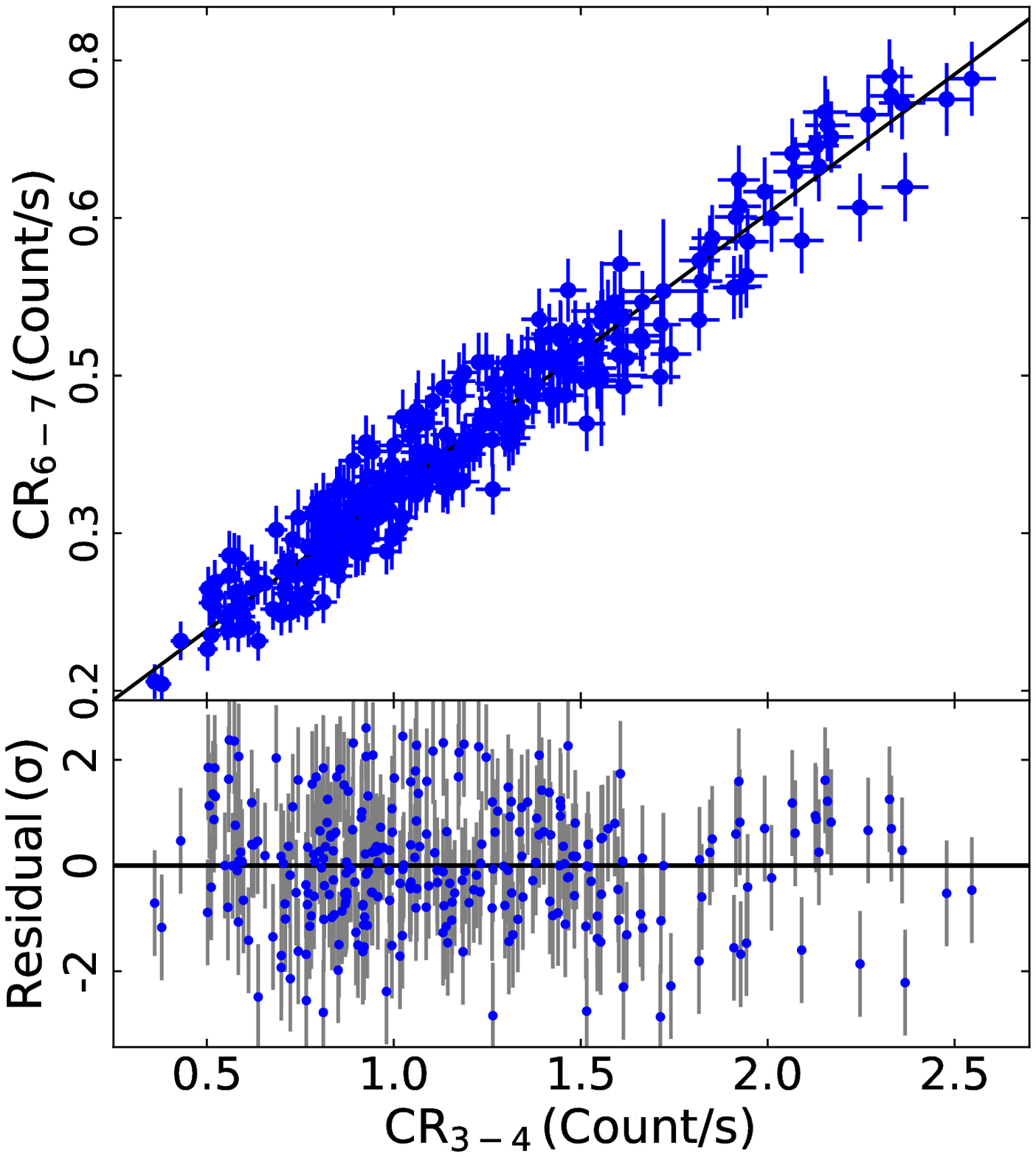}}
{\includegraphics[width=0.225\textwidth]{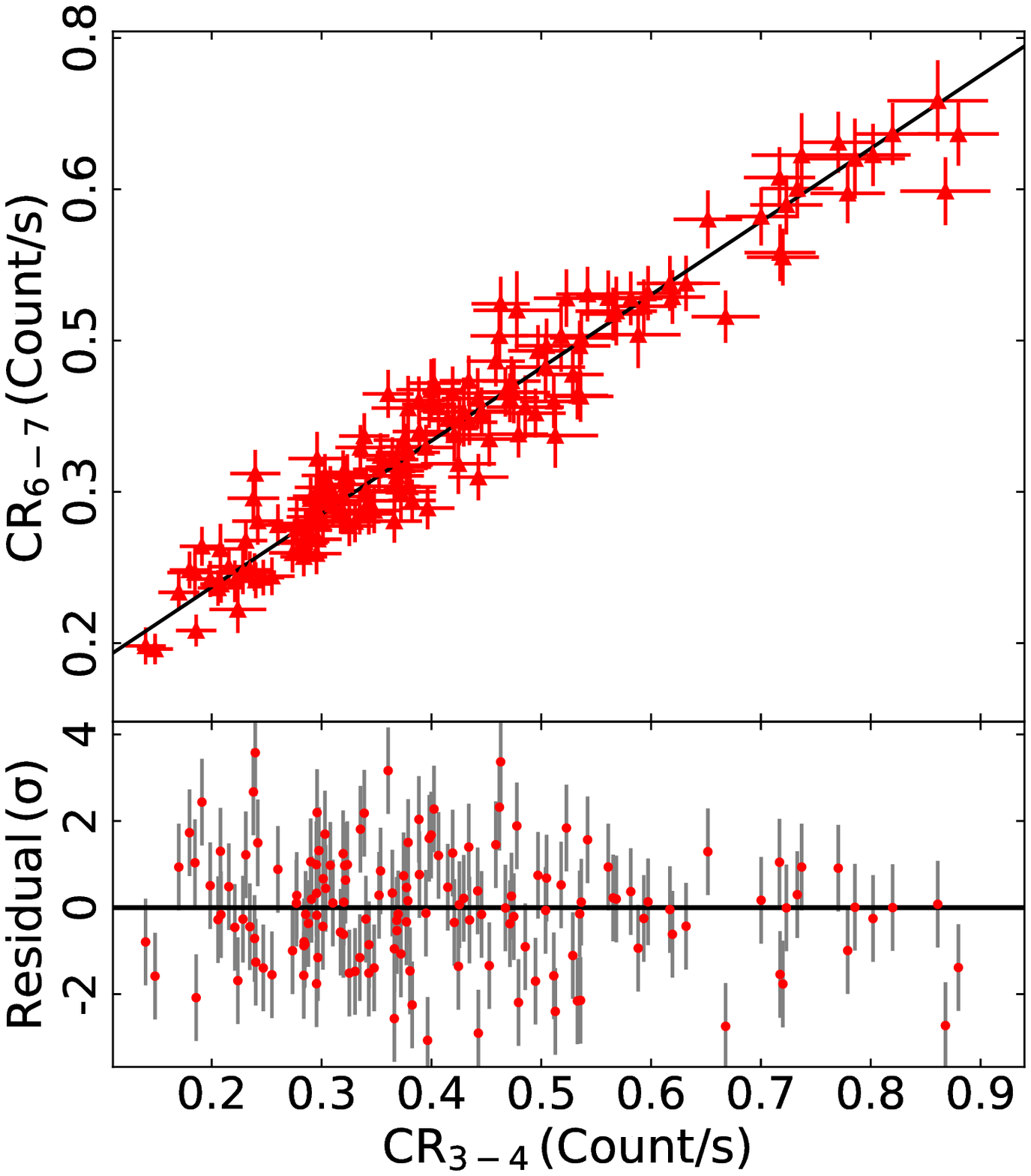}}
{\includegraphics[width=0.225\textwidth]{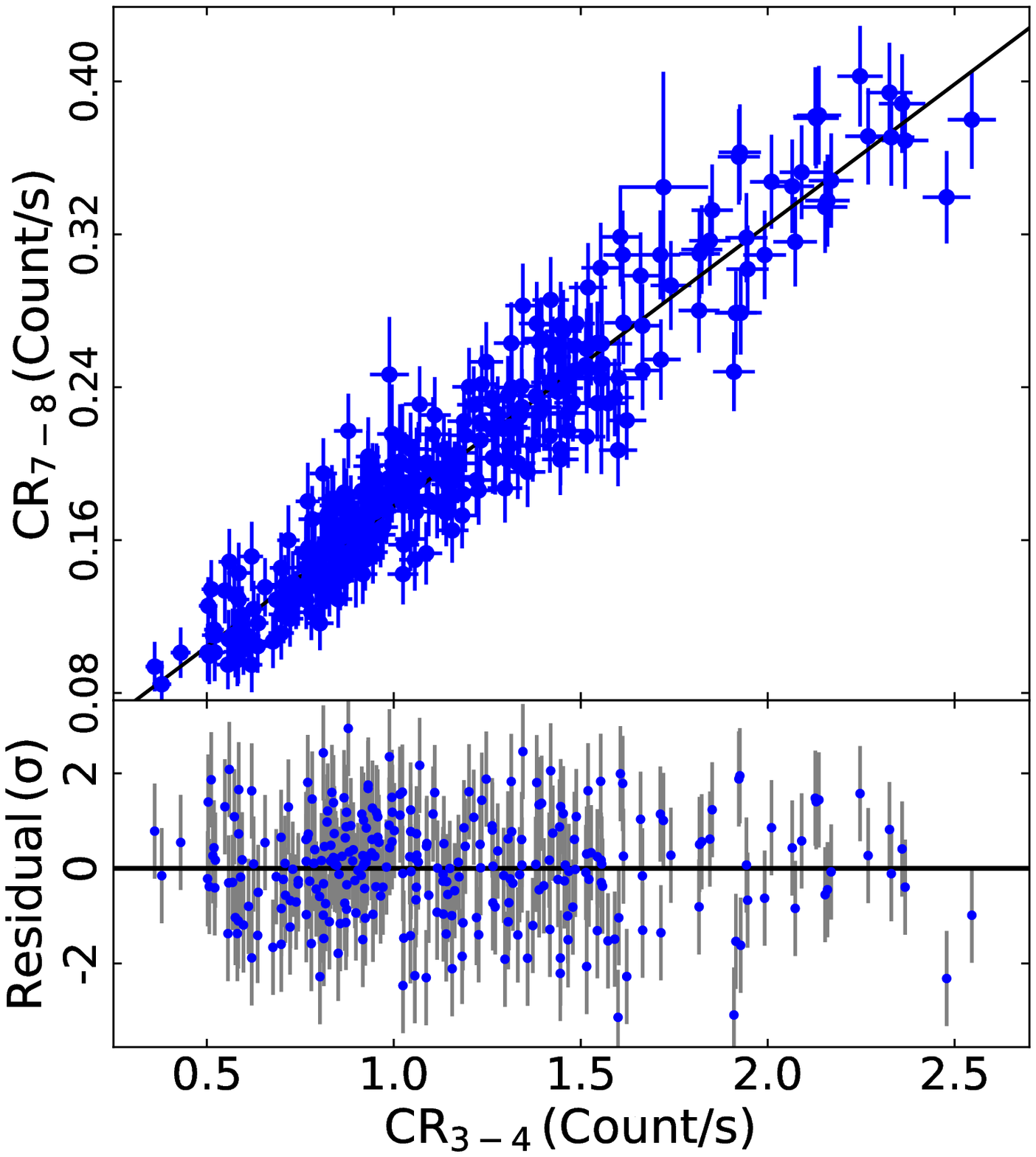}}
{\includegraphics[width=0.225\textwidth]{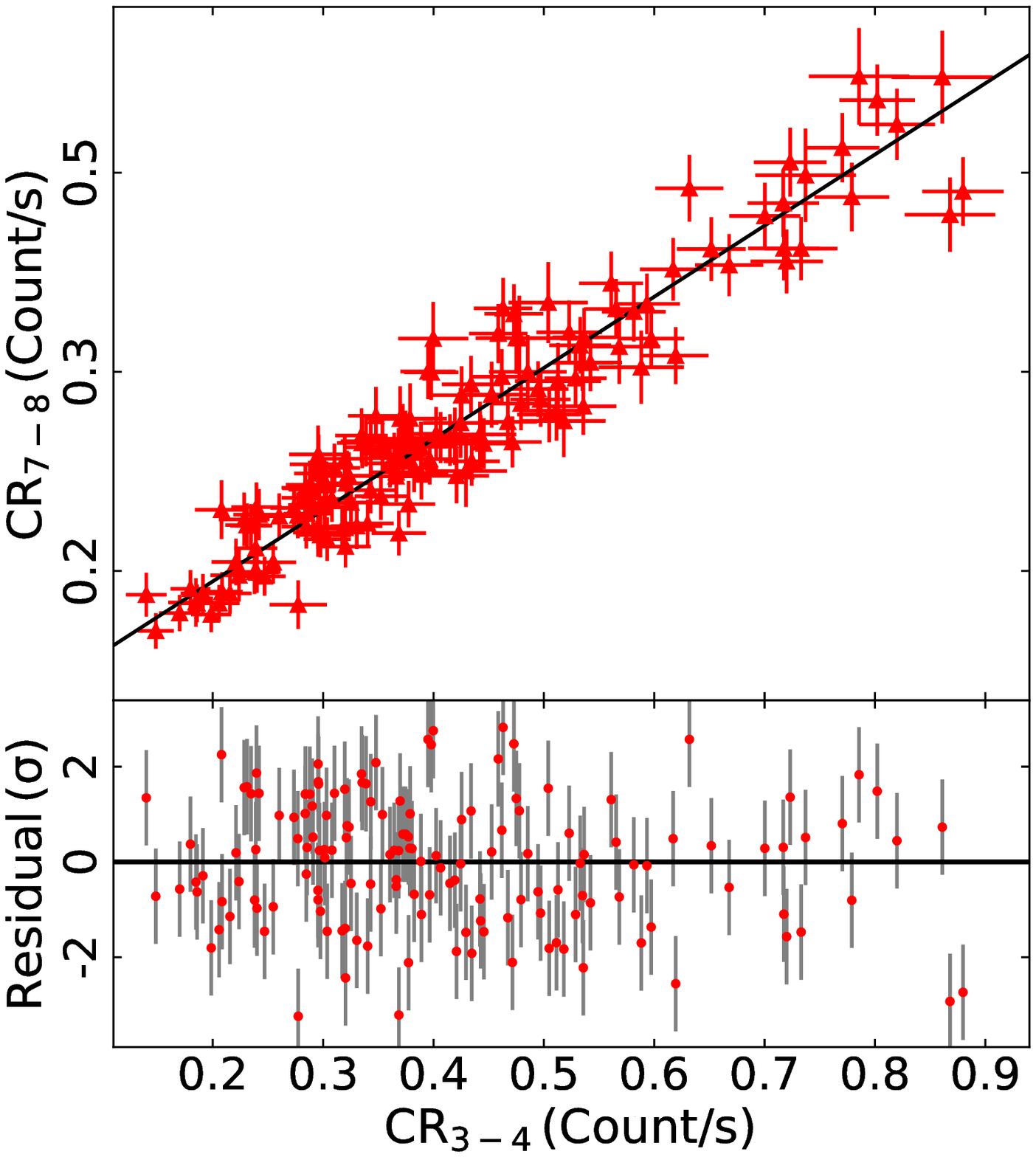}}
{\includegraphics[width=0.225\textwidth]{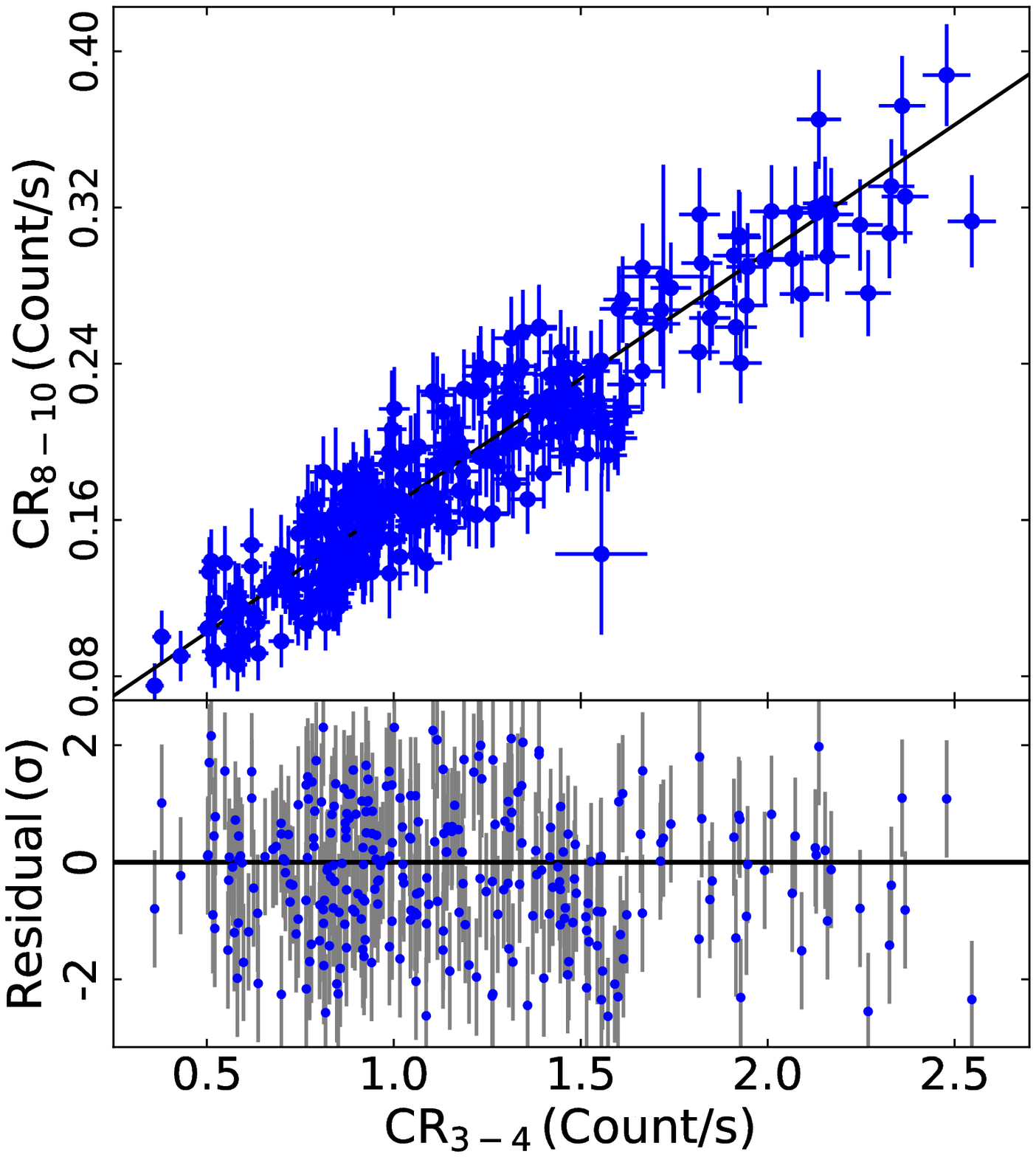}}
{\includegraphics[width=0.225\textwidth]{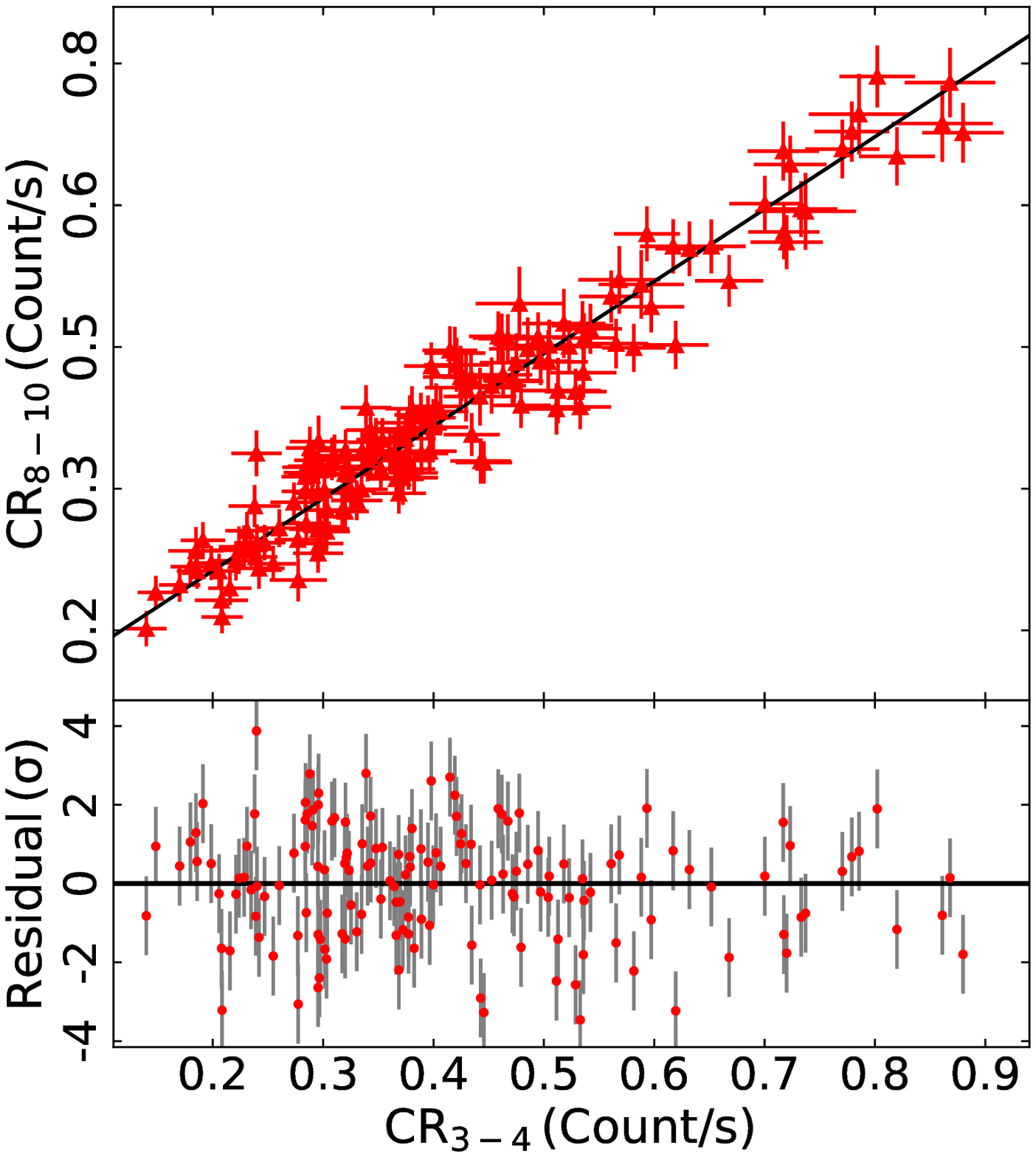}}
\caption{{\it XMM-Newton} and {\it NuSTAR} (left and right column, respectively), high-energy FFPs in the common energy bands (4--10\,keV). The solid black line indicates the best-fit linear model to the combined FFPs. Best-fit residuals are plotted in the lower panel of each plot. }
\label{figapp:commFFPs}
\end{figure}

\newpage

\begin{figure}
\centering

{\includegraphics[width=0.235\textwidth]{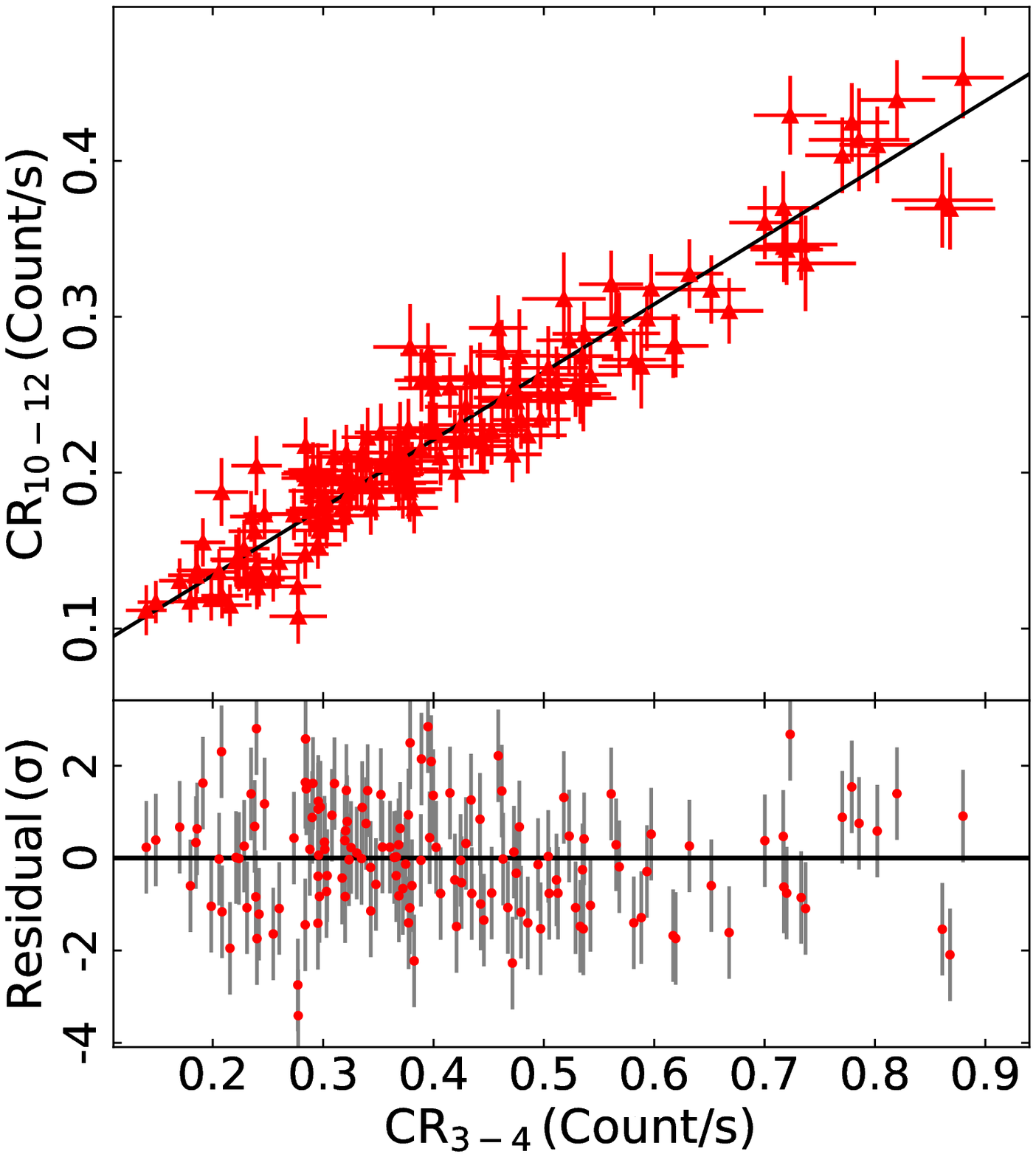}}
{\includegraphics[width=0.235\textwidth]{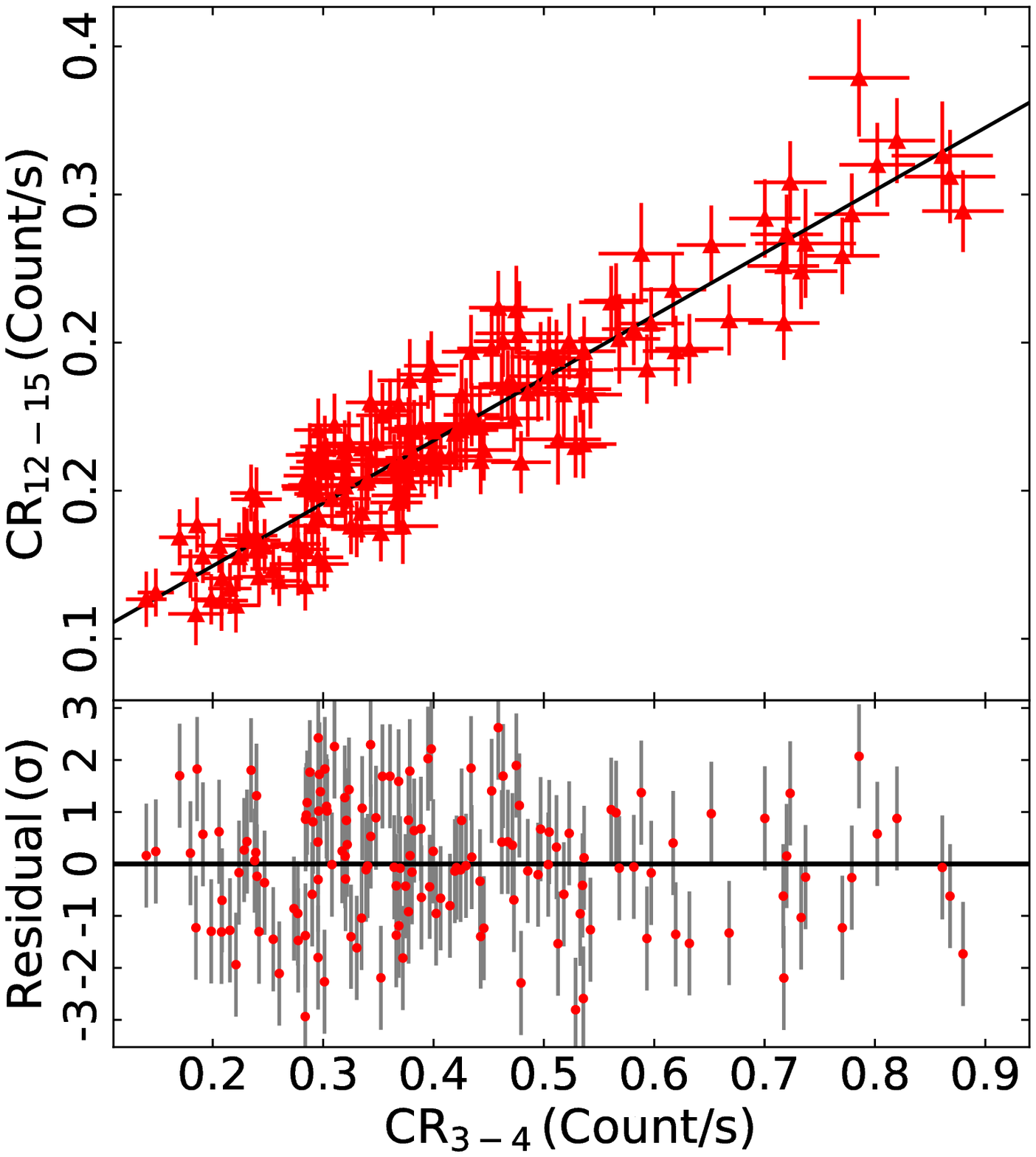}}
{\includegraphics[width=0.235\textwidth]{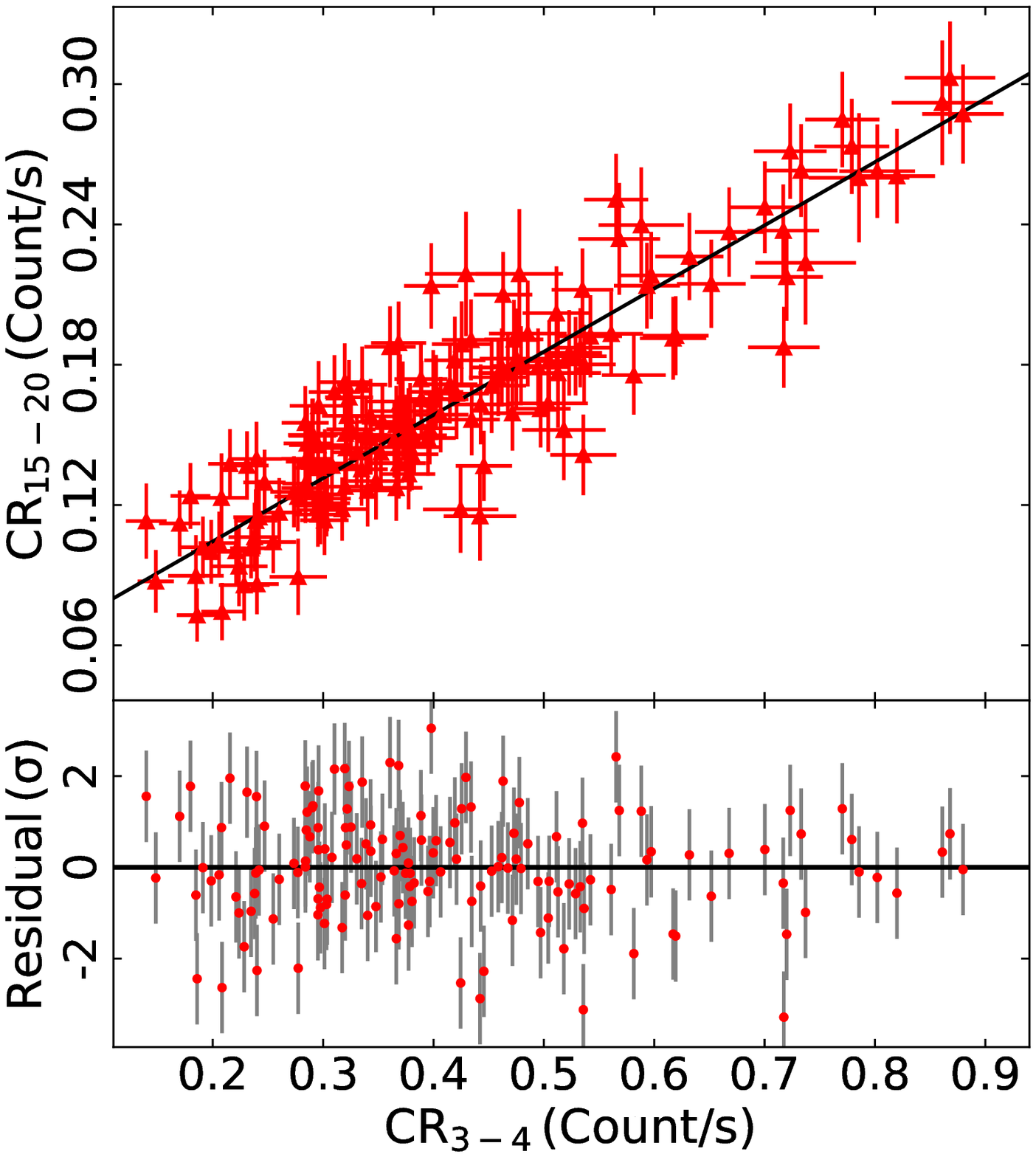}}
{\includegraphics[width=0.235\textwidth]{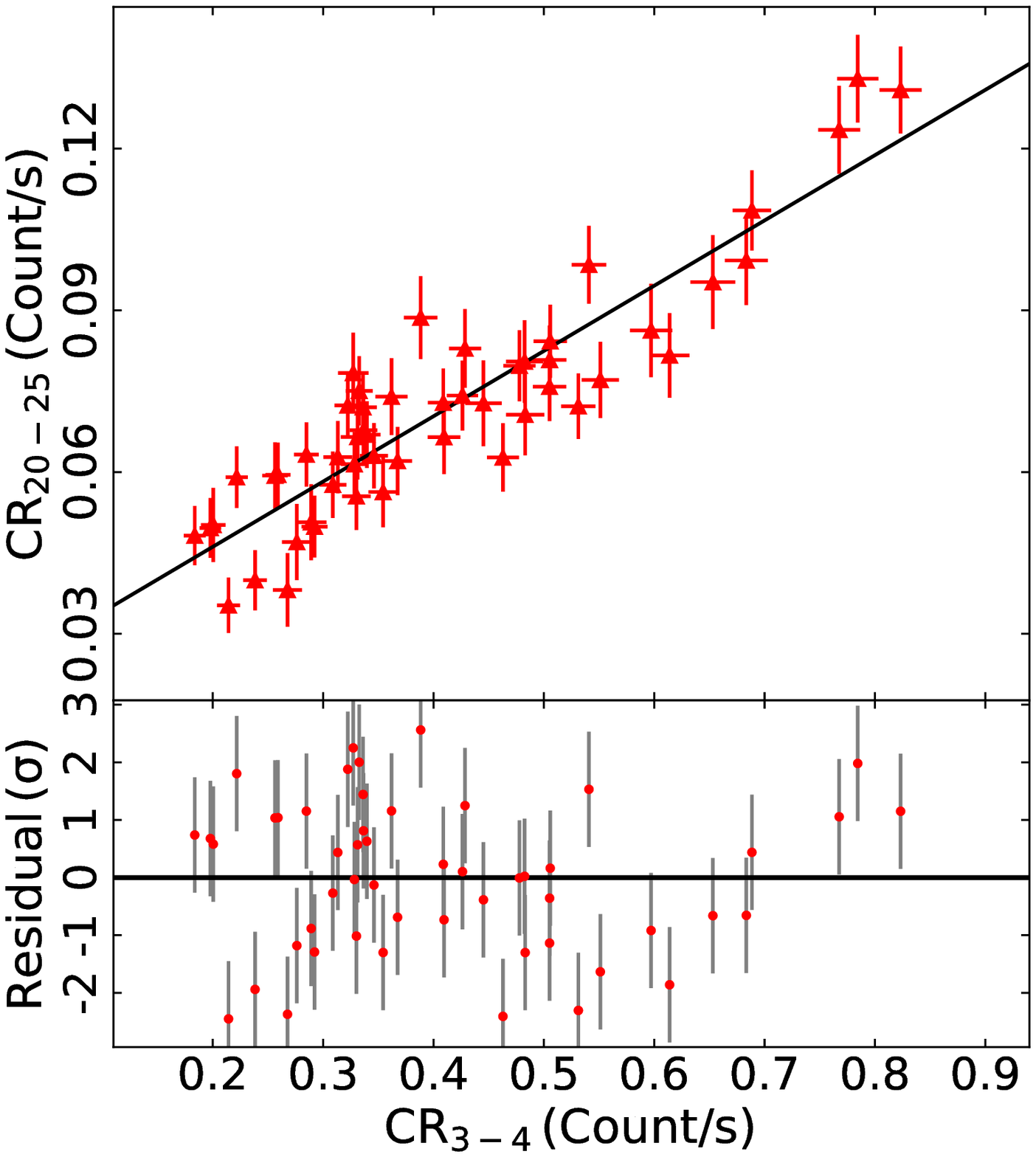}}
{\includegraphics[width=0.235\textwidth]{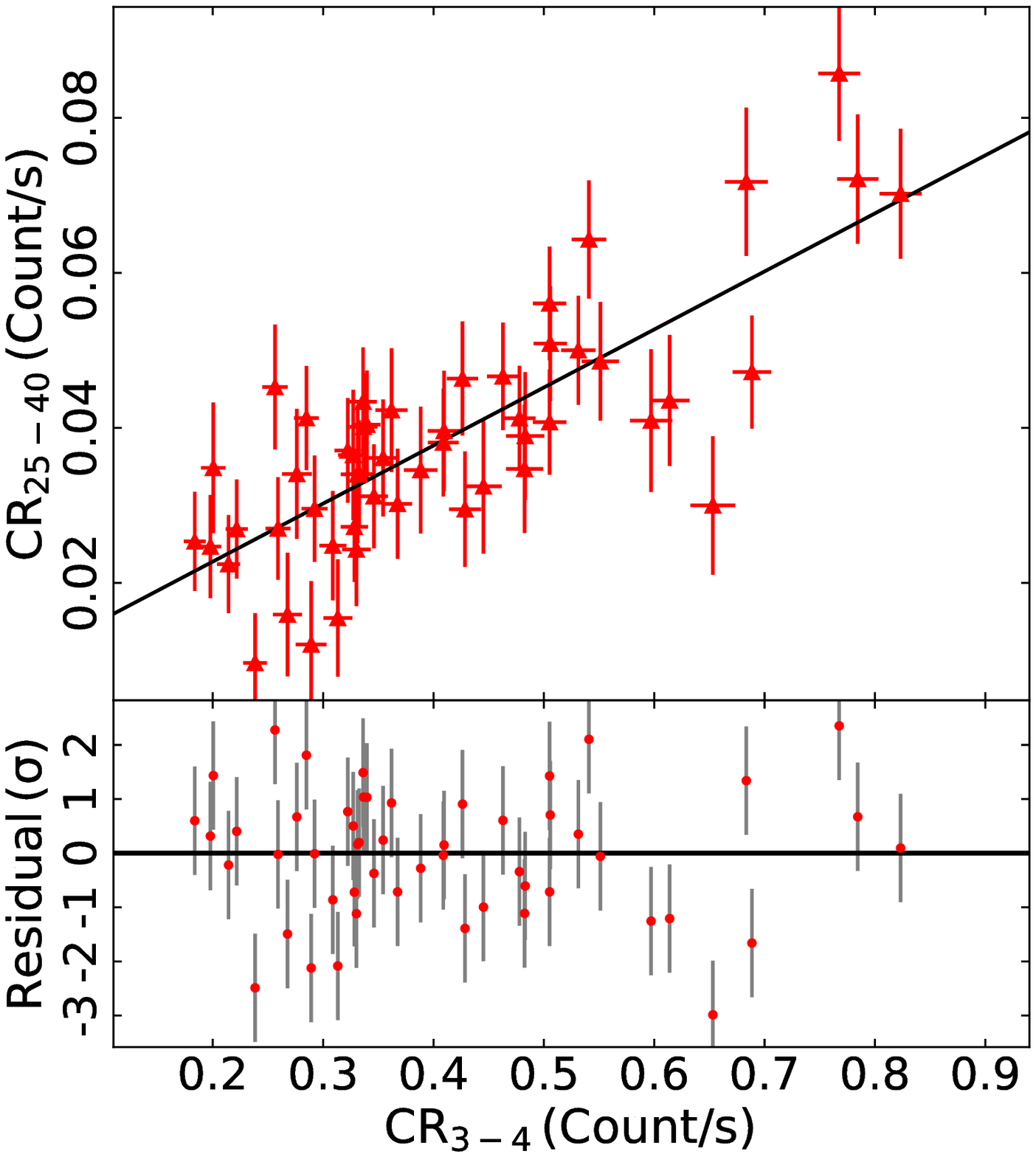}}
\caption{Similar to Fig.\,\ref{figapp:commFFPs} but for the {\it NuSTAR}-only FFPs, in the energy range 10--40\,keV.}
\label{figapp:nustarFFPs}
\end{figure}

\begin{figure*}
\centering

{\includegraphics[scale = 0.25]{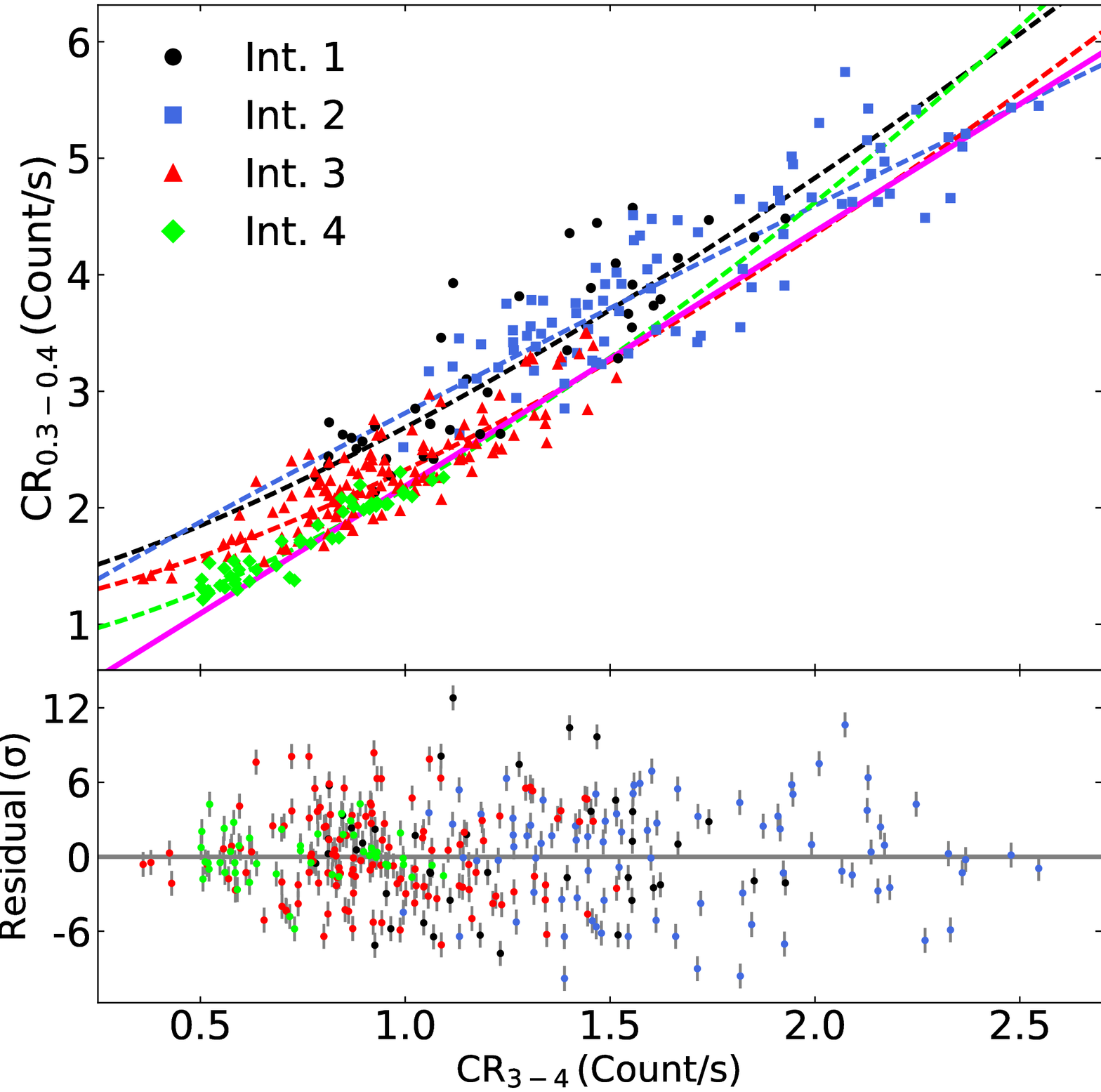}}
{\includegraphics[scale = 0.25]{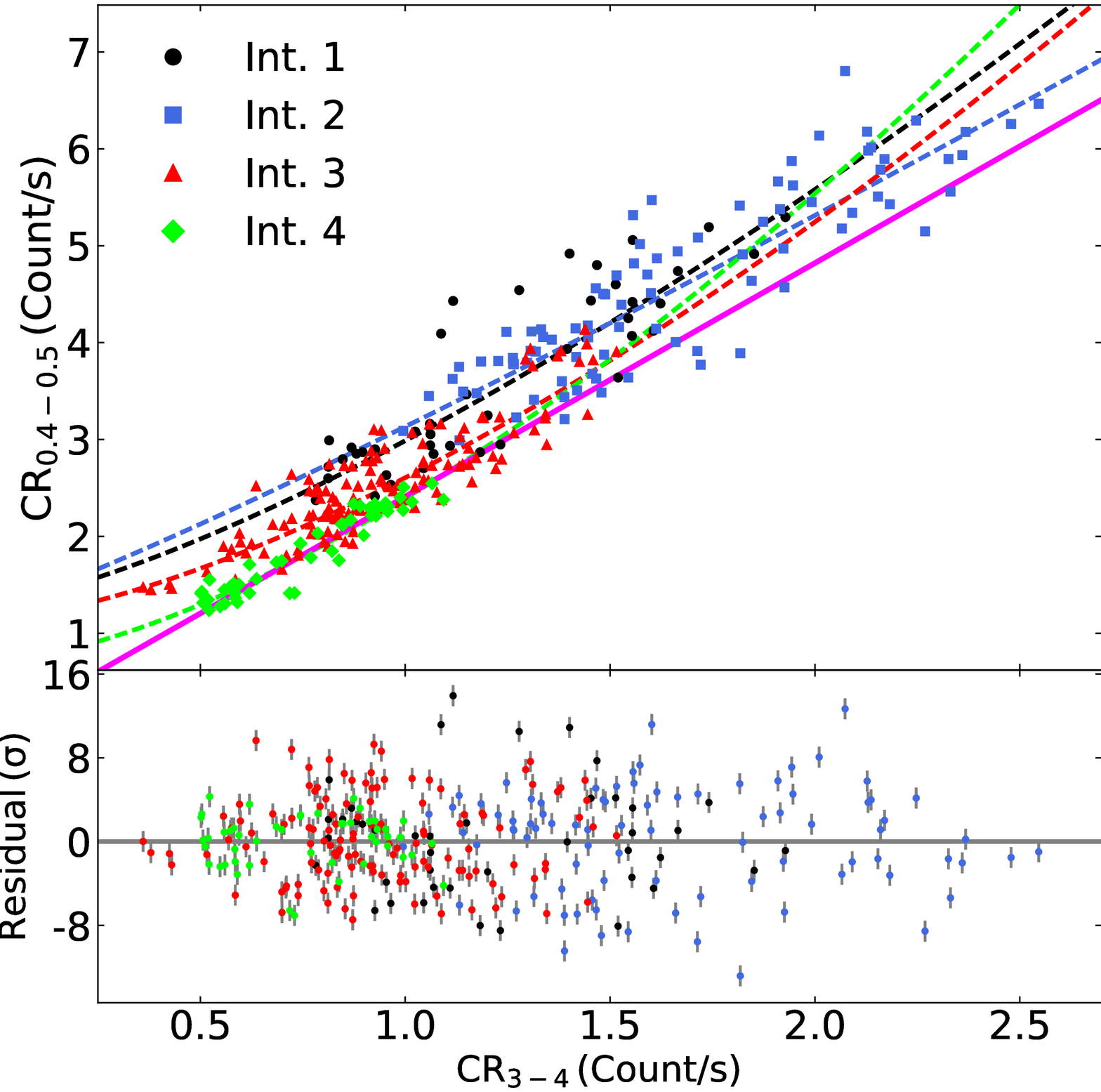}}
{\includegraphics[scale = 0.25]{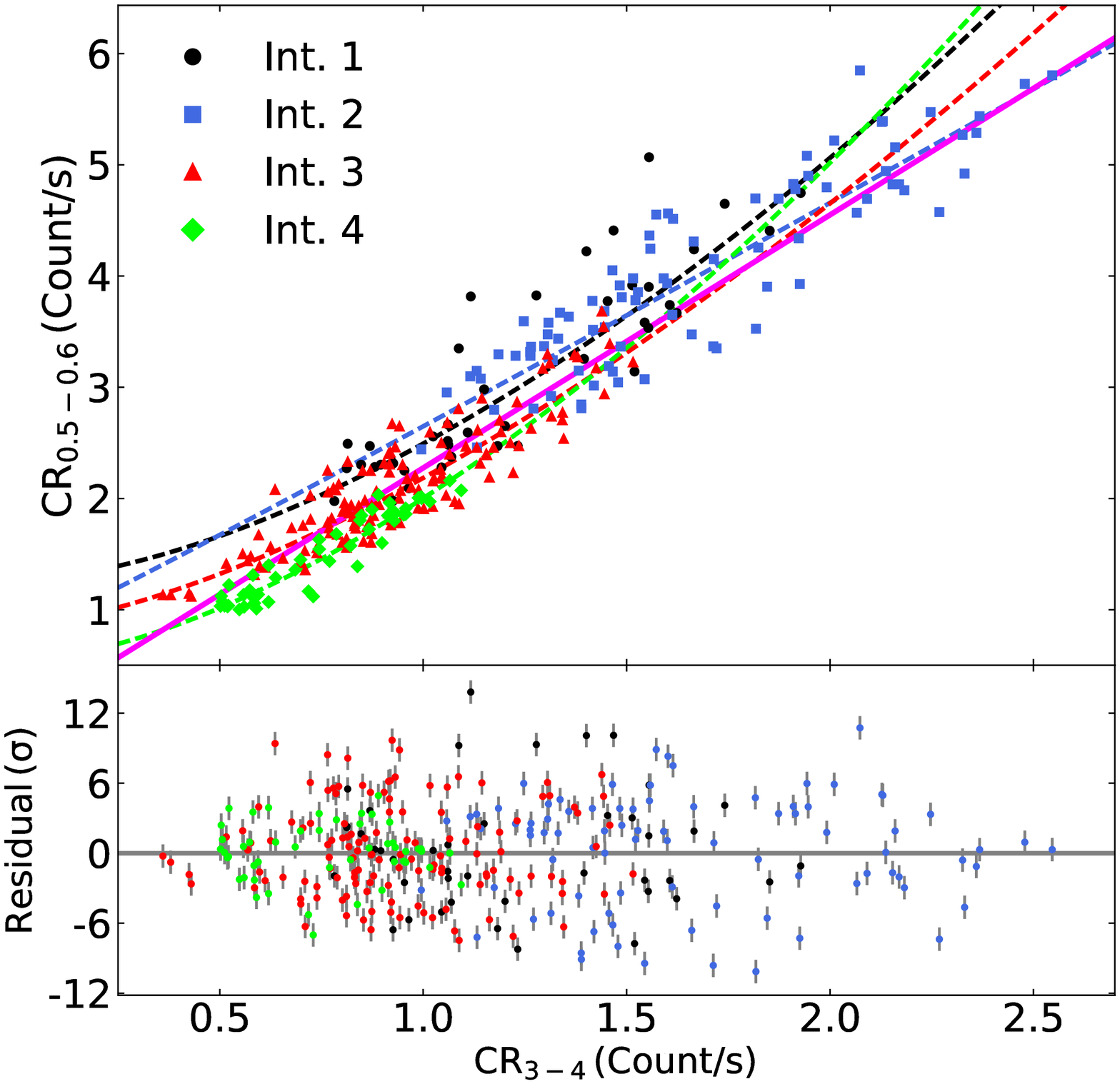}}
{\includegraphics[scale = 0.25]{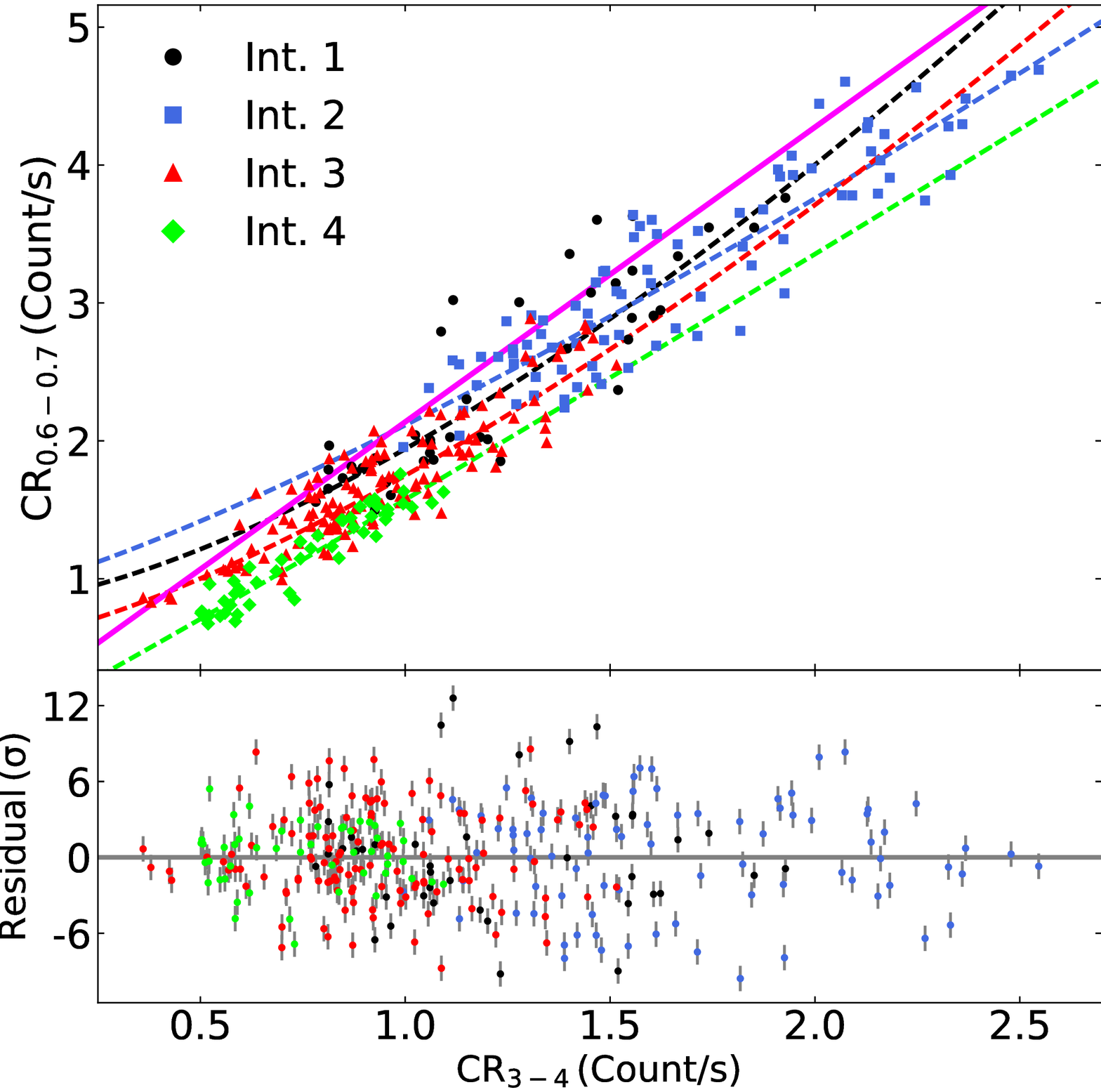}}
{\includegraphics[scale = 0.25]{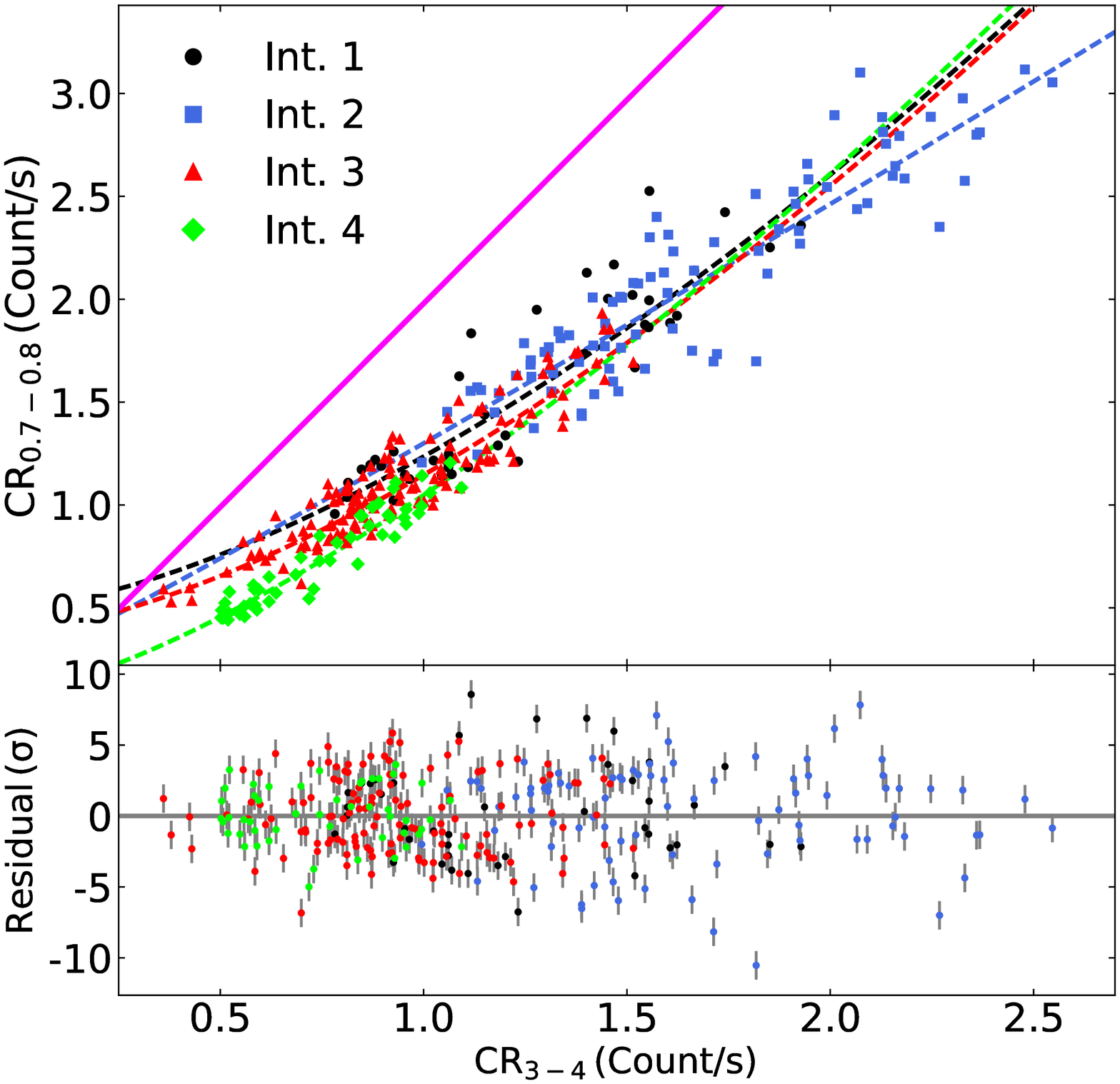}}
{\includegraphics[scale = 0.25]{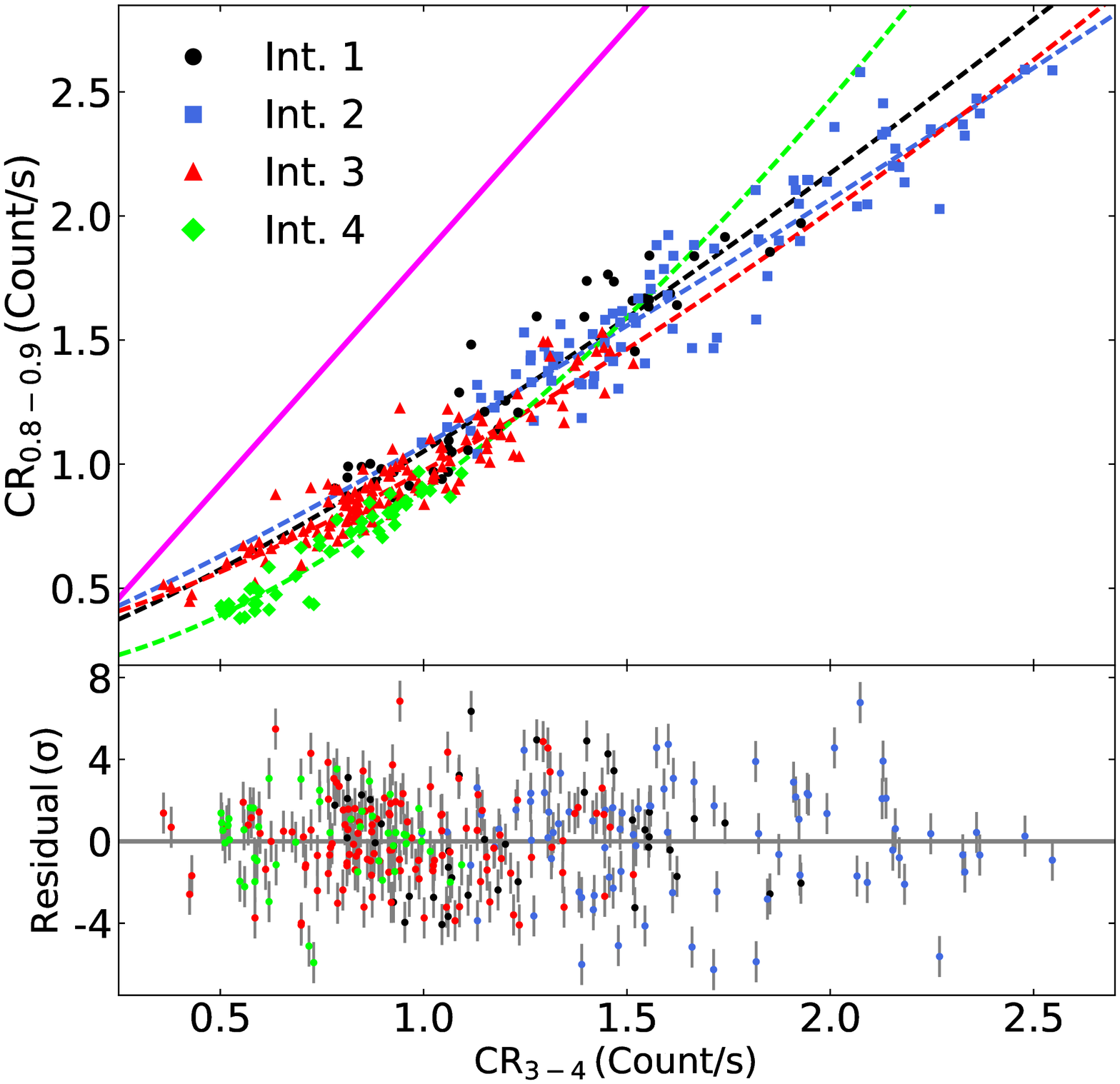}}
{\includegraphics[scale = 0.25]{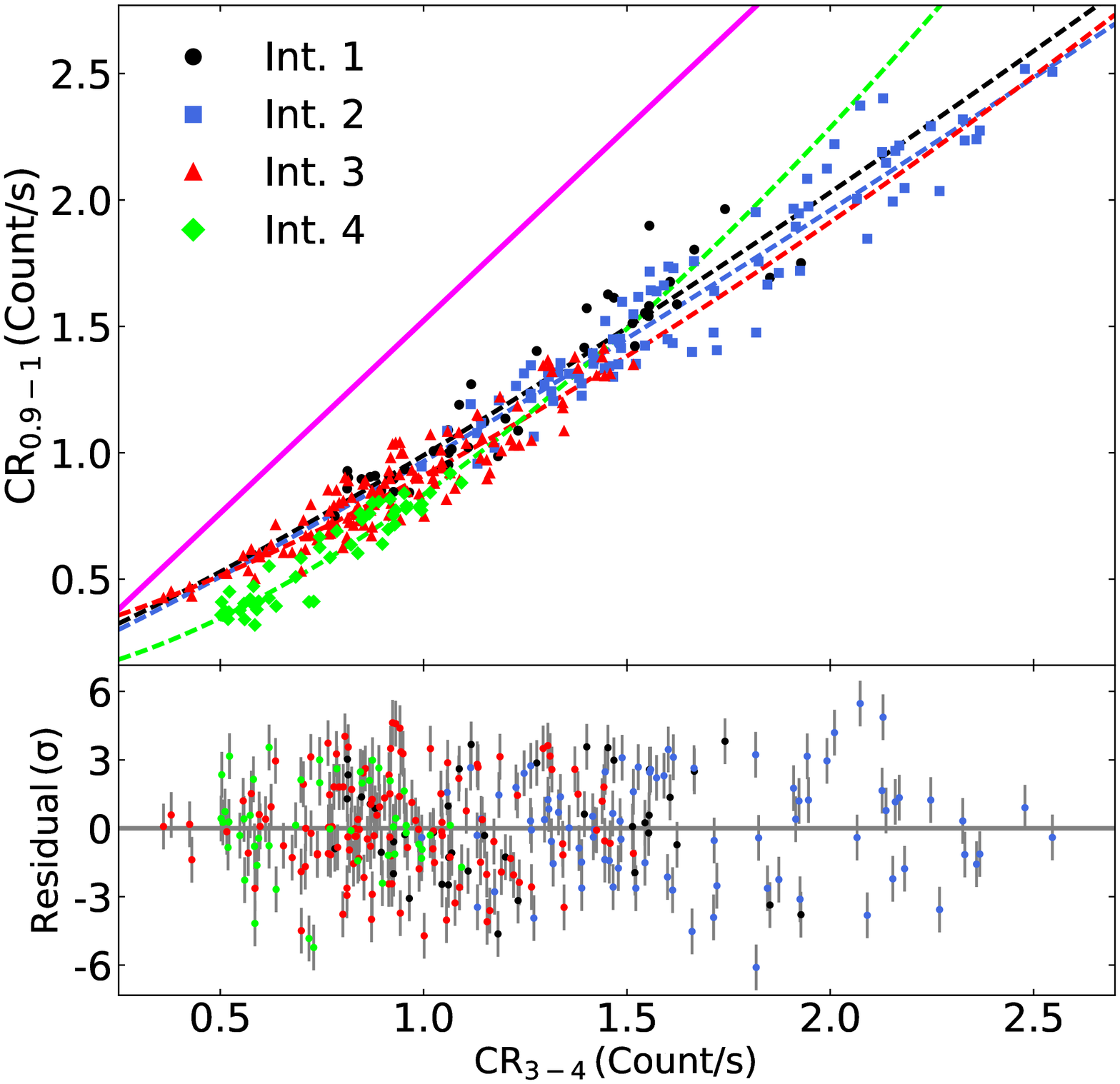}}
{\includegraphics[scale = 0.25]{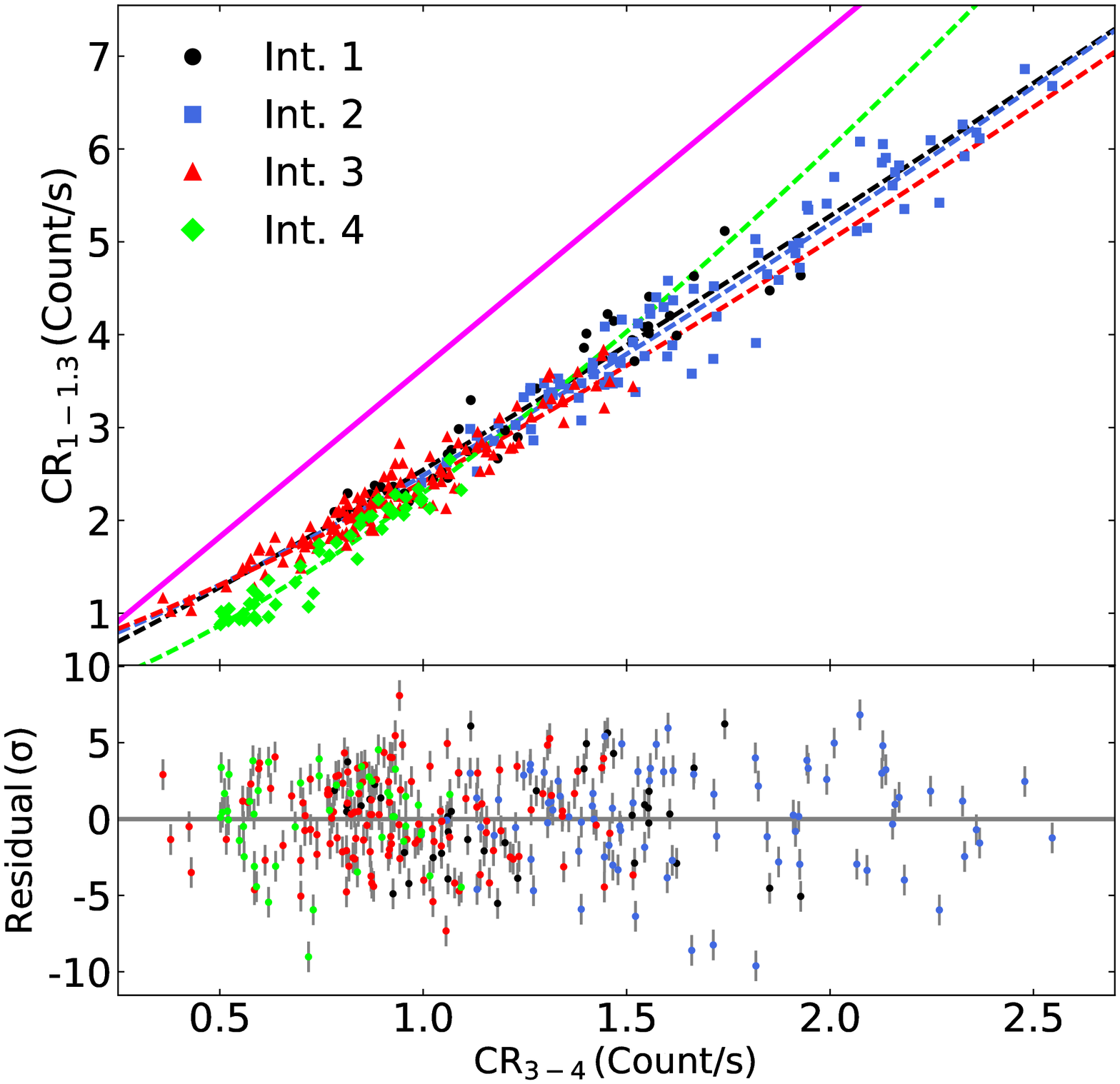}}
{\includegraphics[scale = 0.25]{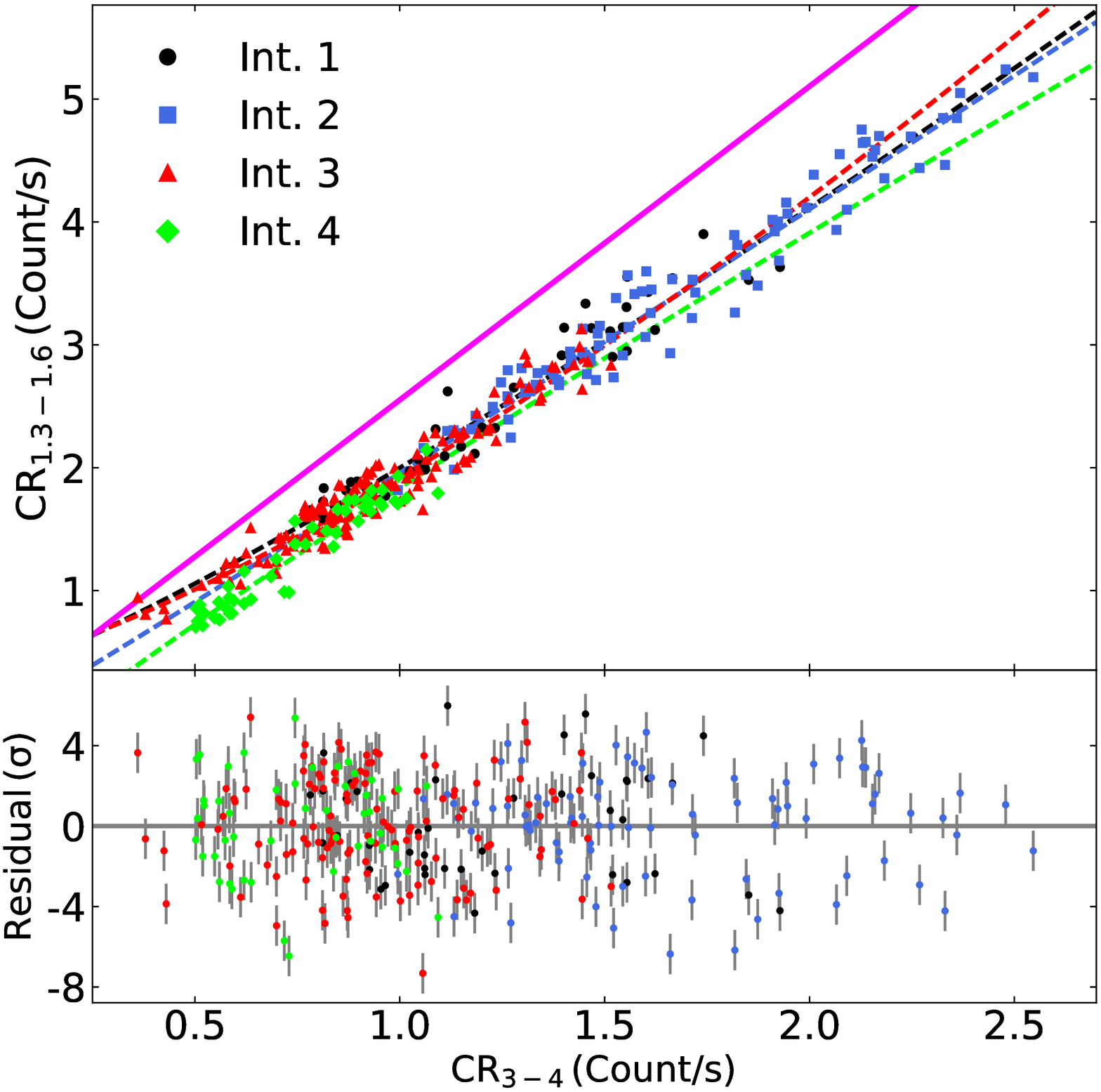}}
{\includegraphics[scale = 0.25]{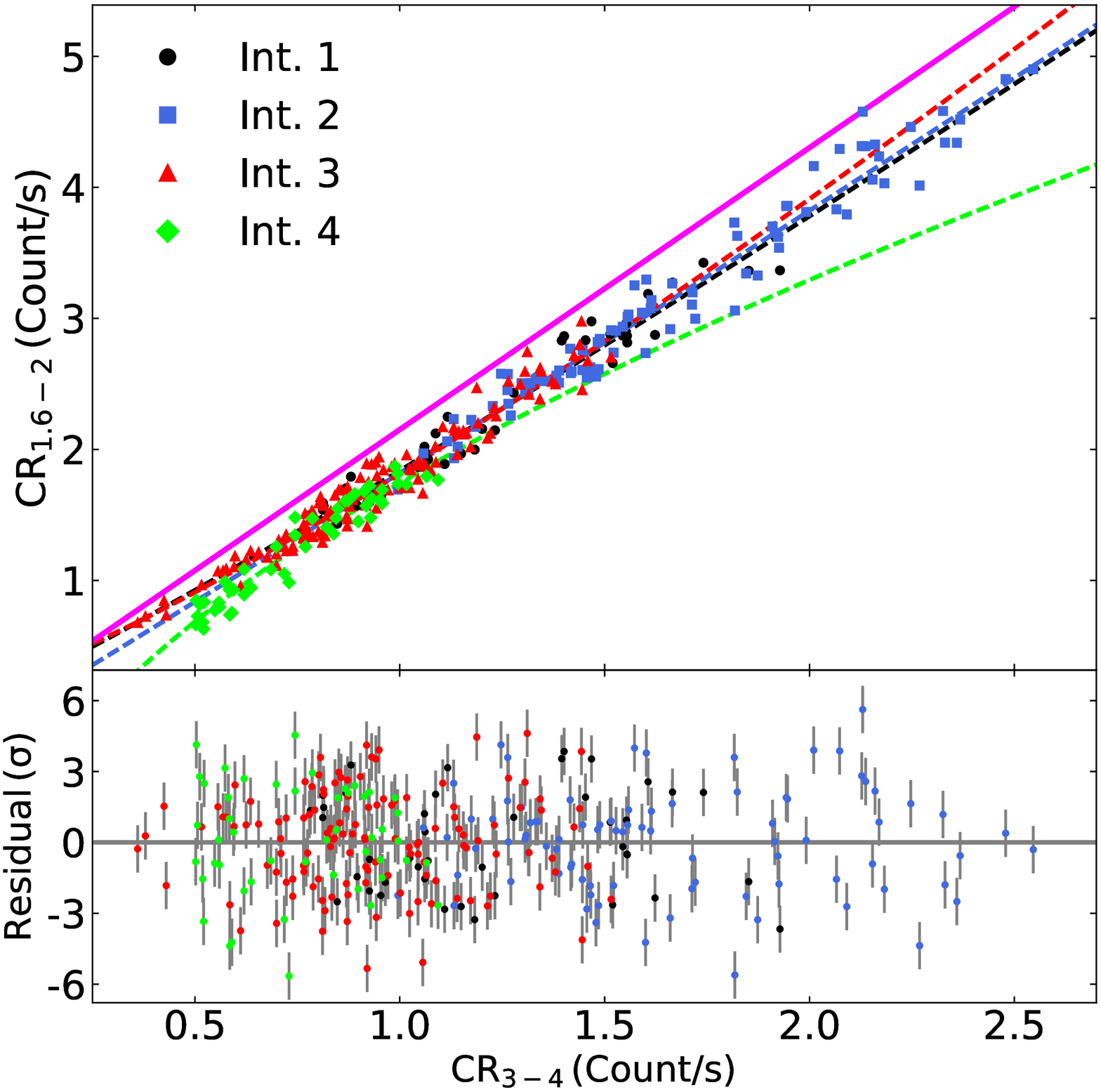}}
{\includegraphics[scale = 0.25]{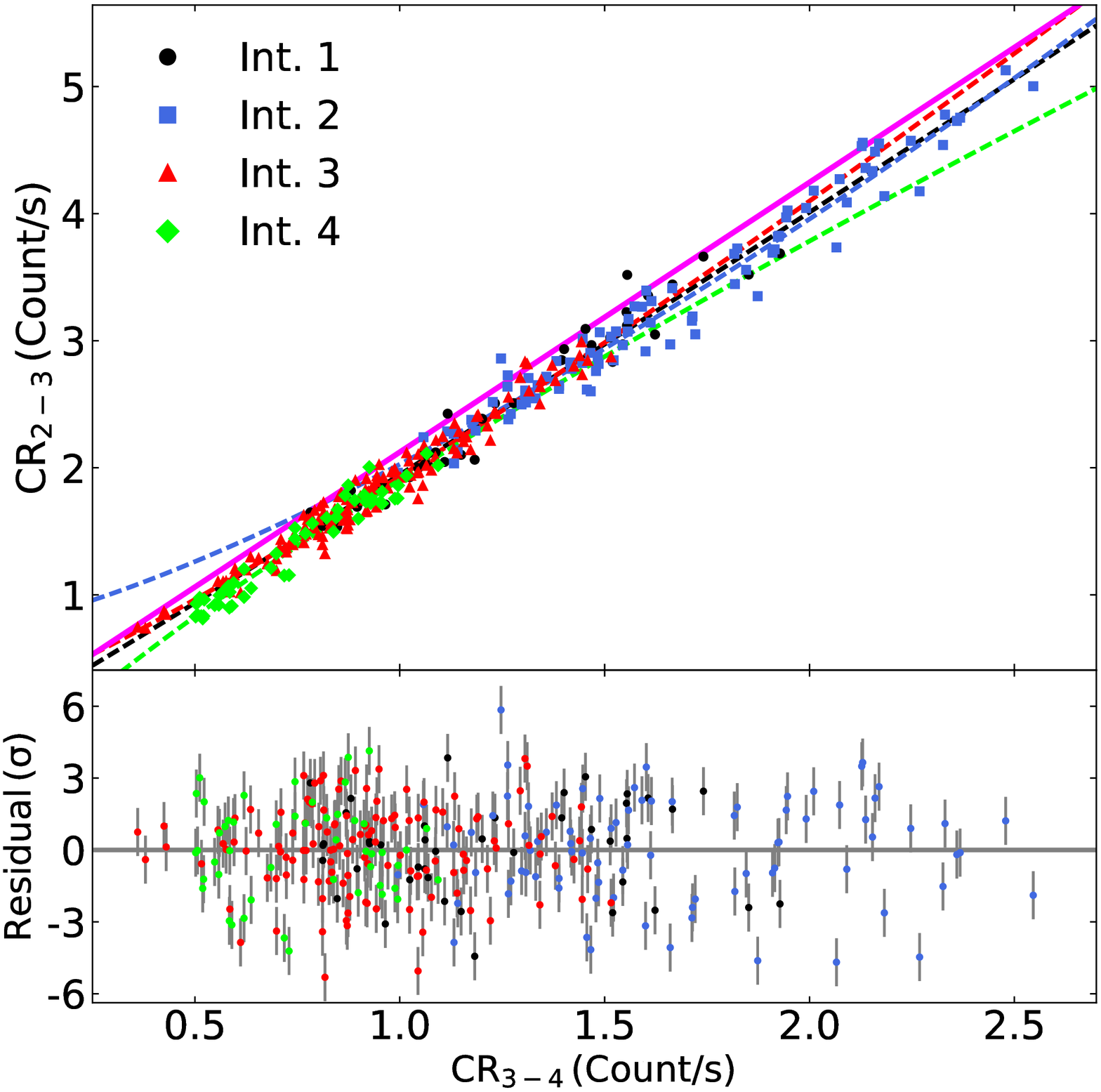}}
\caption{Low-energy FFPs in the 0.3--3\,keV energy range, 
for the individual time intervals Int.\,1-4 (black circles, 
blue squares, red triangles, green diamond, respectively). 
The dashed lines correspond to the best-fit PLc model obtained 
by fitting the data from each time interval separately, using the 
same color code. The solid magenta line indicates the {\it predicted} 
FFPs assuming a power-law spectrum with $\Gamma_{\rm X} = 2.04$ 
(see Section\,\ref{subsec:highE-FFP} for details). We did not plot the 
error bars for clarity reasons. The best-fit residuals are plotted 
in the lower panel of each plot. }
\label{figapp:lowEFFPs}
\end{figure*}

\clearpage

\section{Tables}
\label{app:tables}

\begin{table}
\centering
\caption{Results from the linear model best-fits to the individual and combined {\it XMM-Newton} high-energy FFPs.}
\begin{tabular}{ccccc}
\hline \hline
Energy Band			&	Int.	&	$	A_{\rm L}			$	&	$	C_{\rm L}			$	&	$	\chi^2/{\rm d.o.f.}			$	\\[0.2cm]  	
(keV)			&		&						&	$	({\rm Count\,s^{-1}})			$	&						\\[-0.1cm]     \hline 	
4	--	5	&	1	&	$	0.66	\pm	0.02	$	&	$	0.01	\pm	0.03	$	&	$	43	/	39	$	\\[0.2cm]   	
			&	2	&	$	0.59	\pm	0.01	$	&	$	0.08	\pm	0.02	$	&	$	85	/	82	$	\\[0.2cm]   	
			&	3	&	$	0.65	\pm	0.01	$	&	$	0.01	\pm	0.01	$	&	$	128	/	127	$	\\[0.2cm]   	
			&	4	&	$	0.55	\pm	0.03	$	&	$	0.10	\pm	0.02	$	&	$	64	/	46	$	\\[0.2cm]  \cline{2-5}	\\[-0.2cm]
			&	mean	&	$	0.625	\pm	0.006	$	&	$	0.037	\pm	0.009	$	&	$		-		$	\\[0.2cm] 	
			&	all	&	$	0.622	\pm	0.006	$	&	$	0.044	\pm	0.006	$	&	$	330	/	300	$	\\[0.2cm]   \hline	
5	--	6	&	1	&	$	0.40	\pm	0.02	$	&	$	0.06	\pm	0.02	$	&	$	54	/	39	$	\\[0.2cm]   	
			&	2	&	$	0.39	\pm	0.01	$	&	$	0.08	\pm	0.02	$	&	$	98	/	82	$	\\[0.2cm]   	
			&	3	&	$	0.43	\pm	0.01	$	&	$	0.03	\pm	0.01	$	&	$	115	/	127	$	\\[0.2cm]   	
			&	4	&	$	0.39	\pm	0.02	$	&	$	0.08	\pm	0.02	$	&	$	62	/	46	$	\\[0.2cm]  \cline{2-5}	\\[-0.2cm]
			&	mean	&	$	0.406	\pm	0.005	$	&	$	0.049	\pm	0.007	$	&	$		-		$	\\[0.2cm] 	
			&	all	&	$	0.406	\pm	0.005	$	&	$	0.056	\pm	0.005	$	&	$	337	/	300	$	\\[0.2cm]   \hline	
6	--	7	&	1	&	$	0.24	\pm	0.01	$	&	$	0.11	\pm	0.02	$	&	$	35	/	39	$	\\[0.2cm]   	
			&	2	&	$	0.25	\pm	0.01	$	&	$	0.09	\pm	0.01	$	&	$	108	/	82	$	\\[0.2cm]   	
			&	3	&	$	0.27	\pm	0.01	$	&	$	0.06	\pm	0.01	$	&	$	142	/	127	$	\\[0.2cm]   	
			&	4	&	$	0.23	\pm	0.02	$	&	$	0.10	\pm	0.01	$	&	$	63	/	46	$	\\[0.2cm]  \cline{2-5}	\\[-0.2cm]
			&	mean	&	$	0.257	\pm	0.005	$	&	$	0.079	\pm	0.006	$	&	$		-		$	\\[0.2cm] 	
			&	all	&	$	0.264	\pm	0.004	$	&	$	0.075	\pm	0.004	$	&	$	325	/	300	$	\\[0.2cm]   \hline	
7	--	8	&	1	&	$	0.15	\pm	0.01	$	&	$	0.03	\pm	0.01	$	&	$	49	/	39	$	\\[0.2cm]   	
			&	2	&	$	0.16	\pm	0.01	$	&	$	0.03	\pm	0.01	$	&	$	106	/	82	$	\\[0.2cm]   	
			&	3	&	$	0.15	\pm	0.01	$	&	$	0.02	\pm	0.01	$	&	$	126	/	127	$	\\[0.2cm]   	
			&	4	&	$	0.17	\pm	0.01	$	&	$	0.02	\pm	0.01	$	&	$	40	/	46	$	\\[0.2cm]  \cline{2-5}	\\[-0.2cm]
			&	mean	&	$	0.155	\pm	0.004	$	&	$	0.023	\pm	0.004	$	&	$		-		$	\\[0.2cm] 	
			&	all	&	$	0.147	\pm	0.003	$	&	$	0.031	\pm	0.003	$	&	$	316	/	300	$	\\[0.2cm]   \hline	
8	--	10	&	1	&	$	0.12	\pm	0.01	$	&	$	0.04	\pm	0.01	$	&	$	42	/	39	$	\\[0.2cm]   	
			&	2	&	$	0.13	\pm	0.01	$	&	$	0.03	\pm	0.01	$	&	$	124	/	82	$	\\[0.2cm]   	
			&	3	&	$	0.15	\pm	0.01	$	&	$	0.013	\pm	0.005	$	&	$	188	/	127	$	\\[0.2cm]   	
			&	4	&	$	0.13	\pm	0.01	$	&	$	0.04	\pm	0.01	$	&	$	47	/	46	$	\\[0.2cm]  \cline{2-5}	\\[-0.2cm]
			&	mean	&	$	0.136	\pm	0.004	$	&	$	0.023	\pm	0.004	$	&	$		-		$	\\[0.2cm] 	
			&	all	&	$	0.130	\pm	0.003	$	&	$	0.037	\pm	0.003	$	&	$	344	/	300	$	\\[0.2cm]   \hline \hline		
\end{tabular}
\label{table:XMM-highE}
\end{table}

\begin{table}
\centering
\caption{Similar to Table\,\ref{table:XMM-highE} but for {\it NuSTAR}.}
\begin{tabular}{ccccc}
\hline \hline
Energy Band			&	Int.	&	$	A_{\rm L}			$	&	$	C_{\rm L}			$	&	$	\chi^2/{\rm d.o.f.}			$	\\[0.2cm]  	
(keV)			&		&						&	$	({\rm Count\,s^{-1}})			$	&						\\[-0.1cm]     \hline 	
4	--	5	&	1	&	$	1.04	\pm	0.08	$	&	$	0.01	\pm	0.03	$	&	$	29	/	20	$	\\[0.2cm]   	
			&	2	&	$	1.02	\pm	0.05	$	&	$	0.02	\pm	0.03	$	&	$	40	/	42	$	\\[0.2cm]   	
			&	3	&	$	1.04	\pm	0.05	$	&	$	0.02	\pm	0.01	$	&	$	87	/	68	$	\\[0.2cm]   	
			&	4	&	$	0.93	\pm	0.10	$	&	$	0.05	\pm	0.02	$	&	$	27	/	24	$	\\[0.2cm]  \cline{2-5}	\\[-0.2cm]
			&	mean	&	$	1.025	\pm	0.031	$	&	$	0.021	\pm	0.011	$	&	$		-		$	\\[0.2cm] 	
			&	all	&	$	1.020	\pm	0.020	$	&	$	0.021	\pm	0.007	$	&	$	162	/	160	$	\\[0.2cm]   \hline	
5	--	6	&	1	&	$	0.75	\pm	0.06	$	&	$	0.10	\pm	0.03	$	&	$	17	/	20	$	\\[0.2cm]   	
			&	2	&	$	0.86	\pm	0.05	$	&	$	0.05	\pm	0.03	$	&	$	45	/	42	$	\\[0.2cm]   	
			&	3	&	$	0.88	\pm	0.04	$	&	$	0.04	\pm	0.01	$	&	$	60	/	68	$	\\[0.2cm]   	
			&	4	&	$	0.84	\pm	0.09	$	&	$	0.06	\pm	0.02	$	&	$	31	/	24	$	\\[0.2cm]  \cline{2-5}	\\[-0.2cm]
			&	mean	&	$	0.848	\pm	0.027	$	&	$	0.055	\pm	0.010	$	&	$		-		$	\\[0.2cm] 	
			&	all	&	$	0.863	\pm	0.018	$	&	$	0.052	\pm	0.007	$	&	$	136	/	160	$	\\[0.2cm]   \hline	
6	--	7	&	1	&	$	0.61	\pm	0.06	$	&	$	0.12	\pm	0.02	$	&	$	24	/	20	$	\\[0.2cm]   	
			&	2	&	$	0.71	\pm	0.04	$	&	$	0.06	\pm	0.02	$	&	$	42	/	42	$	\\[0.2cm]   	
			&	3	&	$	0.79	\pm	0.04	$	&	$	0.03	\pm	0.01	$	&	$	72	/	68	$	\\[0.2cm]   	
			&	4	&	$	0.58	\pm	0.08	$	&	$	0.10	\pm	0.02	$	&	$	31	/	24	$	\\[0.2cm]  \cline{2-5}	\\[-0.2cm]
			&	mean	&	$	0.716	\pm	0.024	$	&	$	0.063	\pm	0.009	$	&	$		-		$	\\[0.2cm] 	
			&	all	&	$	0.724	\pm	0.016	$	&	$	0.061	\pm	0.006	$	&	$	162	/	160	$	\\[0.2cm]   \hline	
7	--	8	&	1	&	$	0.53	\pm	0.05	$	&	$	0.03	\pm	0.02	$	&	$	31	/	20	$	\\[0.2cm]   	
			&	2	&	$	0.55	\pm	0.03	$	&	$	0.02	\pm	0.01	$	&	$	64	/	42	$	\\[0.2cm]   	
			&	3	&	$	0.56	\pm	0.03	$	&	$	0.03	\pm	0.01	$	&	$	82	/	68	$	\\[0.2cm]   	
			&	4	&	$	0.67	\pm	0.07	$	&	$	0.001	\pm	0.02	$	&	$	22	/	24	$	\\[0.2cm]  \cline{2-5}	\\[-0.2cm]
			&	mean	&	$	0.563	\pm	0.020	$	&	$	0.023	\pm	0.008	$	&	$		-		$	\\[0.2cm] 	
			&	all	&	$	0.536	\pm	0.013	$	&	$	0.034	\pm	0.005	$	&	$	182	/	160	$	\\[0.2cm]   \hline	
8	--	10	&	1	&	$	0.73	\pm	0.06	$	&	$	0.05	\pm	0.02	$	&	$	44	/	20	$	\\[0.2cm]   	
			&	2	&	$	0.79	\pm	0.04	$	&	$	0.03	\pm	0.02	$	&	$	55	/	42	$	\\[0.2cm]   	
			&	3	&	$	0.87	\pm	0.04	$	&	$	0.02	\pm	0.01	$	&	$	69	/	68	$	\\[0.2cm]   	
			&	4	&	$	0.88	\pm	0.09	$	&	$	0.04	\pm	0.02	$	&	$	35	/	24	$	\\[0.2cm]  \cline{2-5}	\\[-0.2cm]
			&	mean	&	$	0.822	\pm	0.026	$	&	$	0.034	\pm	0.010	$	&	$		-		$	\\[0.2cm] 	
			&	all	&	$	0.765	\pm	0.017	$	&	$	0.060	\pm	0.006	$	&	$	187	/	160	$	\\[0.2cm]   \hline	\hline
\end{tabular}
\label{table:Nustar-highE}
\end{table}
		
\begin{table}
\centering
\contcaption{ }
\begin{tabular}{ccccc}
\hline \hline
Energy Band			&	Int.	&	$	A_{\rm L}			$	&	$	C_{\rm L}			$	&	$	\chi^2/{\rm d.o.f.}			$	\\[0.2cm]  	
(keV)			&		&						&	$	({\rm Count\,s^{-1}})			$	&						\\[-0.1cm]     \hline 			
10	--	12	&	1	&	$	0.36	\pm	0.04	$	&	$	0.07	\pm	0.02	$	&	$	12	/	20	$	\\[0.2cm]   	
			&	2	&	$	0.48	\pm	0.03	$	&	$	0.01	\pm	0.01	$	&	$	43	/	42	$	\\[0.2cm]   	
			&	3	&	$	0.47	\pm	0.03	$	&	$	0.04	\pm	0.01	$	&	$	74	/	68	$	\\[0.2cm]   	
			&	4	&	$	0.53	\pm	0.07	$	&	$	0.03	\pm	0.02	$	&	$	29	/	24	$	\\[0.2cm]  \cline{2-5}	\\[-0.2cm]
			&	mean	&	$	0.462	\pm	0.018	$	&	$	0.036	\pm	0.007	$	&	$		-		$	\\[0.2cm] 	
			&	all	&	$	0.435	\pm	0.012	$	&	$	0.047	\pm	0.005	$	&	$	161	/	160	$	\\[0.2cm]   \hline	
12	--	15	&	1	&	$	0.26	\pm	0.04	$	&	$	0.08	\pm	0.01	$	&	$	24	/	20	$	\\[0.2cm]   	
			&	2	&	$	0.34	\pm	0.02	$	&	$	0.04	\pm	0.02	$	&	$	55	/	42	$	\\[0.2cm]   	
			&	3	&	$	0.35	\pm	0.02	$	&	$	0.040	\pm	0.008	$	&	$	71	/	68	$	\\[0.2cm]   	
			&	4	&	$	0.51	\pm	0.06	$	&	$	0.02	\pm	0.02	$	&	$	18	/	24	$	\\[0.2cm]  \cline{2-5}	\\[-0.2cm]
			&	mean	&	$	0.347	\pm	0.016	$	&	$	0.044	\pm	0.006	$	&	$		-		$	\\[0.2cm] 	
			&	all	&	$	0.338	\pm	0.011	$	&	$	0.052	\pm	0.004	$	&	$	180	/	160	$	\\[0.2cm]   \hline 	
15	--	20	&	1	&	$	0.24	\pm	0.04	$	&	$	0.06	\pm	0.01	$	&	$	33	/	20	$	\\[0.2cm]   	
			&	2	&	$	0.30	\pm	0.02	$	&	$	0.03	\pm	0.01	$	&	$	52	/	42	$	\\[0.2cm]   	
			&	3	&	$	0.34	\pm	0.02	$	&	$	0.025	\pm	0.008	$	&	$	82	/	68	$	\\[0.2cm]   	
			&	4	&	$	0.37	\pm	0.06	$	&	$	0.04	\pm	0.01	$	&	$	26	/	24	$	\\[0.2cm]  \cline{2-5}	\\[-0.2cm]
			&	mean	&	$	0.310	\pm	0.015	$	&	$	0.032	\pm	0.006	$	&	$		-		$	\\[0.2cm] 	
			&	all	&	$	0.270	\pm	0.009	$	&	$	0.051	\pm	0.004	$	&	$	182	/	160	$	\\[0.2cm]   \hline 	
20	--	25	&	1	&	$	0.07	\pm	0.03	$	&	$	0.04	\pm	0.01	$	&	$	8	/	5	$	\\[0.2cm]   	
			&	2	&	$	0.15	\pm	0.02	$	&	$	0.01	\pm	0.01	$	&	$	11	/	12	$	\\[0.2cm]   	
			&	3	&	$	0.13	\pm	0.02	$	&	$	0.020	\pm	0.006	$	&	$	39	/	21	$	\\[0.2cm]   	
			&	4	&	$	0.15	\pm	0.04	$	&	$	0.02	\pm	0.01	$	&	$	2	/	6	$	\\[0.2cm]  \cline{2-5}	\\[-0.2cm]
			&	mean	&	$	0.130	\pm	0.010	$	&	$	0.019	\pm	0.004	$	&	$		-		$	\\[0.2cm] 	
			&	all	&	$	0.120	\pm	0.006	$	&	$	0.022	\pm	0.002	$	&	$	83	/	50	$	\\[0.2cm]   \hline	
25	--	40	&	1	&	$	0.06	\pm	0.03	$	&	$	0.011	\pm	0.01	$	&	$	6	/	5	$	\\[0.2cm]   	
			&	2	&	$	0.09	\pm	0.02	$	&	$	-0.002	\pm	0.01	$	&	$	26	/	12	$	\\[0.2cm]   	
			&	3	&	$	0.11	\pm	0.02	$	&	$	-0.005	\pm	0.006	$	&	$	21	/	21	$	\\[0.2cm]   	
			&	4	&	$	0.07	\pm	0.05	$	&	$	0.02	\pm	0.01	$	&	$	5	/	6	$	\\[0.2cm]  \cline{2-5}	\\[-0.2cm]
			&	mean	&	$	0.090	\pm	0.010	$	&	$	0.019	\pm	0.004	$	&	$		-		$	\\[0.2cm] 	
			&	all	&	$	0.075	\pm	0.006	$	&	$	0.008	\pm	0.003	$	&	$	73	/	50	$	\\[0.2cm]   \hline \hline	

\end{tabular}
\label{table:cont-Nustar-highE}
\end{table}

\begin{table*}
\centering
\caption{The values of best-fit parameters obtained by fitting the {\it XMM-Newton} low-energy FFPs with a PLc model (eq.\,\ref{eq:PLc}) for the individual time intervals and their arithmetic mean. In addition we show the best-fit results obtained by fitting the data from the 4 time intervals together.}
\begin{tabular}{cccccc}
\hline \hline
Energy Band			&	Int.	&	$	A_{\rm PLc}			$	&	$	\beta$			&	$	C_{\rm PLc}			$	&	$	\chi^2/{\rm d.o.f.}			$	\\[0.2cm]  	
(keV)			&		&						&	$				$	&	$	({\rm Count\,s^{-1}})			$	&						\\[-0.1cm]     \hline 	
0.3	--	0.4	&	1	&	$	1.39	\pm	0.21	$	&	$	1.34	\pm	0.15	$	&	$	1.30	\pm	0.21	$	&	$	968	/	40	$	\\[0.2cm]   	
			&	2	&	$	1.97	\pm	0.36	$	&	$	0.93	\pm	0.11	$	&	$	0.84	\pm	0.39	$	&	$	1634	/	83	$	\\[0.2cm]   	
			&	3	&	$	1.18	\pm	0.06	$	&	$	1.44	\pm	0.07	$	&	$	1.14	\pm	0.05	$	&	$	1651	/	128	$	\\[0.2cm]   	
			&	4	&	$	1.36	\pm	0.15	$	&	$	1.49	\pm	0.27	$	&	$	0.79	\pm	0.15	$	&	$	188	/	47	$	\\[0.2cm]  \cline{2-6}	\\[-0.2cm]
			&	mean	&	$	1.47	\pm	0.17	$	&	$	1.30	\pm	0.13	$	&	$	1.02	\pm	0.12	$	&	$		-		$	\\[0.2cm] 	
			&	all	&	$	2.17	\pm	0.04	$	&	$	1.01	\pm	0.02	$	&	$	0.25	\pm	0.04	$	&	$	6674	/	304	$	\\[0.2cm]   \hline	
0.4	--	0.5	&	1	&	$	1.67	\pm	0.22	$	&	$	1.35	\pm	0.14	$	&	$	1.32	\pm	0.22	$	&	$	1166	/	40	$	\\[0.2cm]   	
			&	2	&	$	1.860	\pm	0.27	$	&	$	1.11	\pm	0.10	$	&	$	1.26	\pm	0.29	$	&	$	2151	/	83	$	\\[0.2cm]   	
			&	3	&	$	1.45	\pm	0.05	$	&	$	1.49	\pm	0.06	$	&	$	1.15	\pm	0.05	$	&	$	2156	/	128	$	\\[0.2cm]   	
			&	4	&	$	1.68	\pm	0.14	$	&	$	1.52	\pm	0.22	$	&	$	0.71	\pm	0.15	$	&	$	296	/	47	$	\\[0.2cm]  \cline{2-6}	\\[-0.2cm]
			&	mean	&	$	1.66	\pm	0.08	$	&	$	1.37	\pm	0.09	$	&	$	1.11	\pm	0.14	$	&	$		-		$	\\[0.2cm] 	
			&	all	&	$	2.63	\pm	0.05	$	&	$	1.01	\pm	0.02	$	&	$	0.07	\pm	0.04	$	&	$	8119	/	304	$	\\[0.2cm]   \hline	
0.5	--	0.6	&	1	&	$	1.23	\pm	0.14	$	&	$	1.63	\pm	0.14	$	&	$	1.26	\pm	0.14	$	&	$	1080	/	40	$	\\[0.2cm]   	
			&	2	&	$	1.90	\pm	0.29	$	&	$	1.04	\pm	0.10	$	&	$	0.75	\pm	0.31	$	&	$	1906	/	83	$	\\[0.2cm]   	
			&	3	&	$	1.33	\pm	0.05	$	&	$	1.52	\pm	0.06	$	&	$	0.86	\pm	0.05	$	&	$	2063	/	128	$	\\[0.2cm]   	
			&	4	&	$	1.47	\pm	0.12	$	&	$	1.61	\pm	0.23	$	&	$	0.54	\pm	0.12	$	&	$	304	/	47	$	\\[0.2cm]  \cline{2-6}	\\[-0.2cm]
			&	mean	&	$	1.48	\pm	0.15	$	&	$	1.45	\pm	0.14	$	&	$	0.85	\pm	0.15	$	&	$		-		$	\\[0.2cm] 	
			&	all	&	$	2.33	\pm	0.04	$	&	$	1.04	\pm	0.02	$	&	$	-0.07	\pm	0.04	$	&	$	7435	/	304	$	\\[0.2cm]   \hline	
0.6	--	0.7	&	1	&	$	1.12	\pm	0.15	$	&	$	1.50	\pm	0.15	$	&	$	0.82	\pm	0.15	$	&	$	988	/	40	$	\\[0.2cm]   	
			&	2	&	$	1.21	\pm	0.18	$	&	$	1.24	\pm	0.11	$	&	$	0.91	\pm	0.20	$	&	$	1524	/	83	$	\\[0.2cm]   	
			&	3	&	$	1.21	\pm	0.05	$	&	$	1.40	\pm	0.06	$	&	$	0.54	\pm	0.05	$	&	$	1728	/	128	$	\\[0.2cm]   	
			&	4	&	$	1.68	\pm	0.25	$	&	$	1.04	\pm	0.22	$	&	$	-0.11	\pm	0.26	$	&	$	293	/	47	$	\\[0.2cm]  \cline{2-6}	\\[-0.2cm]
			&	mean	&	$	1.31	\pm	0.13	$	&	$	1.30	\pm	0.10	$	&	$	0.54	\pm	0.23	$	&	$		-		$	\\[0.2cm] 	
			&	all	&	$	1.99	\pm	0.04	$	&	$	1.01	\pm	0.02	$	&	$	-0.20	\pm	0.04	$	&	$	6072	/	304	$	\\[0.2cm]   \hline	
0.7	--	0.8	&	1	&	$	0.73	\pm	0.12	$	&	$	1.52	\pm	0.17	$	&	$	0.50	\pm	0.11	$	&	$	489	/	40	$	\\[0.2cm]   	
			&	2	&	$	1.08	\pm	0.20	$	&	$	1.06	\pm	0.13	$	&	$	0.22	\pm	0.22	$	&	$	1036	/	83	$	\\[0.2cm]   	
			&	3	&	$	0.76	\pm	0.03	$	&	$	1.52	\pm	0.08	$	&	$	0.39	\pm	0.03	$	&	$	918	/	128	$	\\[0.2cm]   	
			&	4	&	$	0.96	\pm	0.11	$	&	$	1.40	\pm	0.26	$	&	$	0.09	\pm	0.11	$	&	$	191	/	47	$	\\[0.2cm]  \cline{2-6}	\\[-0.2cm]
			&	mean	&	$	0.88	\pm	0.08	$	&	$	1.37	\pm	0.11	$	&	$	0.30	\pm	0.09	$	&	$		-		$	\\[0.2cm] 	
			&	all	&	$	1.29	\pm	0.03	$	&	$	1.02	\pm	0.02	$	&	$	-0.12	\pm	0.03	$	&	$	3435	/	304	$	\\[0.2cm]   \hline \hline	
\end{tabular}
\label{table:XMM-lowE}
\end{table*}

\begin{table*}
\centering
\contcaption{}
\begin{tabular}{cccccc}
\hline \hline
Energy Band			&	Int.	&	$	A_{\rm PLc}			$	&	$	\beta			$	&	$	C_{\rm PLc}			$	&	$	\chi^2/{\rm d.o.f.}			$	\\[0.2cm]  	
(keV)			&		&						&	$				$	&	$	({\rm Count\,s^{-1}})			$	&						\\[-0.1cm]    \hline 	
0.8	--	0.9	&	1	&	$	0.83	\pm	0.16	$	&	$	1.24	\pm	0.19	$	&	$	0.23	\pm	0.16	$	&	$	283	/	40	$	\\[0.2cm]   	
			&	2	&	$	0.81	\pm	0.15	$	&	$	1.16	\pm	0.13	$	&	$	0.27	\pm	0.17	$	&	$	608	/	83	$	\\[0.2cm]   	
			&	3	&	$	0.67	\pm	0.04	$	&	$	1.36	\pm	0.08	$	&	$	0.31	\pm	0.04	$	&	$	602	/	128	$	\\[0.2cm]   	
			&	4	&	$	0.74	\pm	0.07	$	&	$	1.65	\pm	0.29	$	&	$	0.16	\pm	0.08	$	&	$	175	/	47	$	\\[0.2cm]  \cline{2-6}	\\[-0.2cm]
			&	mean	&	$	0.76	\pm	0.04	$	&	$	1.35	\pm	0.11	$	&	$	0.24	\pm	0.03	$	&	$		-		$	\\[0.2cm] 	
			&	all	&	$	1.05	\pm	0.03	$	&	$	1.04	\pm	0.02	$	&	$	-0.07	\pm	0.03	$	&	$	2415	/	304	$	\\[0.2cm]   \hline	
0.9	--	1	&	1	&	$	0.83	\pm	0.17	$	&	$	1.17	\pm	0.19	$	&	$	0.16	\pm	0.17	$	&	$	211	/	40	$	\\[0.2cm]   	
			&	2	&	$	0.84	\pm	0.15	$	&	$	1.12	\pm	0.13	$	&	$	0.12	\pm	0.17	$	&	$	452	/	83	$	\\[0.2cm]   	
			&	3	&	$	0.65	\pm	0.04	$	&	$	1.34	\pm	0.08	$	&	$	0.26	\pm	0.03	$	&	$	610	/	128	$	\\[0.2cm]   	
			&	4	&	$	0.74	\pm	0.08	$	&	$	1.57	\pm	0.29	$	&	$	0.10	\pm	0.08	$	&	$	198	/	47	$	\\[0.2cm]  \cline{2-6}	\\[-0.2cm]
			&	mean	&	$	0.77	\pm	0.04	$	&	$	1.30	\pm	0.10	$	&	$	0.16	\pm	0.03	$	&	$		-		$	\\[0.2cm] 	
			&	all	&	$	1.00	\pm	0.03	$	&	$	1.05	\pm	0.02	$	&	$	-0.08	\pm	0.02	$	&	$	2127	/	304	$	\\[0.2cm]   \hline	
1	--	1.3	&	1	&	$	2.35	\pm	0.31	$	&	$	1.11	\pm	0.12	$	&	$	0.19	\pm	0.30	$	&	$	409	/	40	$	\\[0.2cm]   	
			&	2	&	$	2.08	\pm	0.21	$	&	$	1.20	\pm	0.08	$	&	$	0.40	\pm	0.24	$	&	$	946	/	83	$	\\[0.2cm]   	
			&	3	&	$	1.94	\pm	0.07	$	&	$	1.23	\pm	0.05	$	&	$	0.48	\pm	0.06	$	&	$	1018	/	128	$	\\[0.2cm]   	
			&	4	&	$	2.31	\pm	0.16	$	&	$	1.38	\pm	0.16	$	&	$	-0.02	\pm	0.17	$	&	$	412	/	47	$	\\[0.2cm]  \cline{2-6}	\\[-0.2cm]
			&	mean	&	$	2.17	\pm	0.10	$	&	$	1.23	\pm	0.06	$	&	$	0.26	\pm	0.11	$	&	$		-		$	\\[0.2cm] 	
			&	all	&	$	2.68	\pm	0.04	$	&	$	1.04	\pm	0.01	$	&	$	-0.27	\pm	0.04	$	&	$	4015	/	304	$	\\[0.2cm]   \hline	
1.3	--	1.6	&	1	&	$	1.70	\pm	0.25	$	&	$	1.17	\pm	0.13	$	&	$	0.30	\pm	0.24	$	&	$	296	/	40	$	\\[0.2cm]   	
			&	2	&	$	2.06	\pm	0.26	$	&	$	1.03	\pm	0.09	$	&	$	-0.10	\pm	0.29	$	&	$	543	/	83	$	\\[0.2cm]   	
			&	3	&	$	1.51	\pm	0.05	$	&	$	1.33	\pm	0.05	$	&	$	0.41	\pm	0.05	$	&	$	797	/	128	$	\\[0.2cm]   	
			&	4	&	$	2.40	\pm	0.36	$	&	$	0.90	\pm	0.19	$	&	$	-0.57	\pm	0.37	$	&	$	304	/	47	$	\\[0.2cm]  \cline{2-6}	\\[-0.2cm]
			&	mean	&	$	1.92	\pm	0.20	$	&	$	1.11	\pm	0.09	$	&	$	0.01	\pm	0.22	$	&	$		-		$	\\[0.2cm] 	
			&	all	&	$	2.11	\pm	0.04	$	&	$	1.03	\pm	0.02	$	&	$	-0.19	\pm	0.04	$	&	$	2447	/	304	$	\\[0.2cm]   \hline	
1.6	--	2	&	1	&	$	1.72	\pm	0.27	$	&	$	1.09	\pm	0.14	$	&	$	0.11	\pm	0.27	$	&	$	189	/	40	$	\\[0.2cm]   	
			&	2	&	$	1.93	\pm	0.26	$	&	$	1.03	\pm	0.09	$	&	$	-0.11	\pm	0.28	$	&	$	393	/	83	$	\\[0.2cm]   	
			&	3	&	$	1.60	\pm	0.06	$	&	$	1.21	\pm	0.05	$	&	$	0.22	\pm	0.06	$	&	$	549	/	128	$	\\[0.2cm]   	
			&	4	&	$	3.32	\pm	0.95	$	&	$	0.55	\pm	0.19	$	&	$	-1.58	\pm	0.95	$	&	$	255	/	47	$	\\[0.2cm]  \cline{2-6}	\\[-0.2cm]
			&	mean	&	$	2.14	\pm	0.40	$	&	$	0.97	\pm	0.14	$	&	$	-0.34	\pm	0.42	$	&	$		-		$	\\[0.2cm] 	
			&	all	&	$	2.04	\pm	0.04	$	&	$	0.99	\pm	0.02	$	&	$	-0.23	\pm	0.04	$	&	$	1631	/	304	$	\\[0.2cm]   \hline	
2	--	3	&	1	&	$	1.97	\pm	0.31	$	&	$	1.03	\pm	0.13	$	&	$	-0.02	\pm	0.31	$	&	$	149	/	40	$	\\[0.2cm]   	
			&	2	&	$	1.27	\pm	0.15	$	&	$	1.33	\pm	0.09	$	&	$	0.75	\pm	0.17	$	&	$	384	/	83	$	\\[0.2cm]   	
			&	3	&	$	1.74	\pm	0.06	$	&	$	1.16	\pm	0.05	$	&	$	0.18	\pm	0.06	$	&	$	420	/	128	$	\\[0.2cm]   	
			&	4	&	$	2.51	\pm	0.95	$	&	$	0.55	\pm	0.19	$	&	$	-1.58	\pm	0.95	$	&	$	255	/	47	$	\\[0.2cm]  \cline{2-6}	\\[-0.2cm]
			&	mean	&	$	1.88	\pm	0.26	$	&	$	1.08	\pm	0.11	$	&	$	0.08	\pm	0.28	$	&	$		-		$	\\[0.2cm] 	
			&	all	&	$	2.08	\pm	0.04	$	&	$	0.99	\pm	0.02	$	&	$	-0.15	\pm	0.04	$	&	$	1170	/	304	$	\\[0.2cm]   \hline \hline	
\end{tabular}
\end{table*}

\clearpage

\section{The effects of the warm absorber to the FFPs}
\label{app:warmabs}

In order to investigate the effect of a variable warm absorber on the low-energy FFPs, 
we created simulated spectra using the XSPEC command {\tt FAKEIT}, 
and the EPIC-pn responses, assuming the following model (in {\tt XSPEC} terminology): 
\begin{equation}
{\tt model = TBabs \times zxipcf \times powerlaw}, 
\end{equation}
\noindent
where {\tt TBabs} \citep{tbabs} and {\tt zxipcf} \citep{zxipcf} 
account for the Galactic and the warm asborption, respectively. 
{\tt powerlaw} varied in normalization ($N_{\rm PL}$) only, with $\Gamma$ fixed at 2.03. $N_{\rm PL}$ varied between $N_{\rm PL,min}$ and $N_{\rm PL, max}$, so that the respective model 3--4 keV model count rate were equal to the minimum/maximum observed count rate in the same band. As for {\tt zxipcf}, we fixed $N_{\rm H}$ at $2 \times 10^{22}\,{\rm cm^{-2}}$ and we considered 3 different values for the covering fraction (CF): 0.4, 0.6 and 0.8. We assumed that the ionization parameter ($\xi$) is linearly proportional to the primary flux, as: $\log \xi = \log N_{\rm PL} + 2.97$. The constant was chosen so that the model count rate in the 0.6--0.7 keV band (when CF=0.6 and $N_{\rm PL}=N_{\rm PL,max}$) is equal to the observed largest value. Given the $N_{\rm PL,min} -N_{\rm PL, max}$ range, $\log \xi$ varied between 0.85 and 1.55.

\begin{figure}

\centering
\includegraphics[width = 0.49\textwidth]{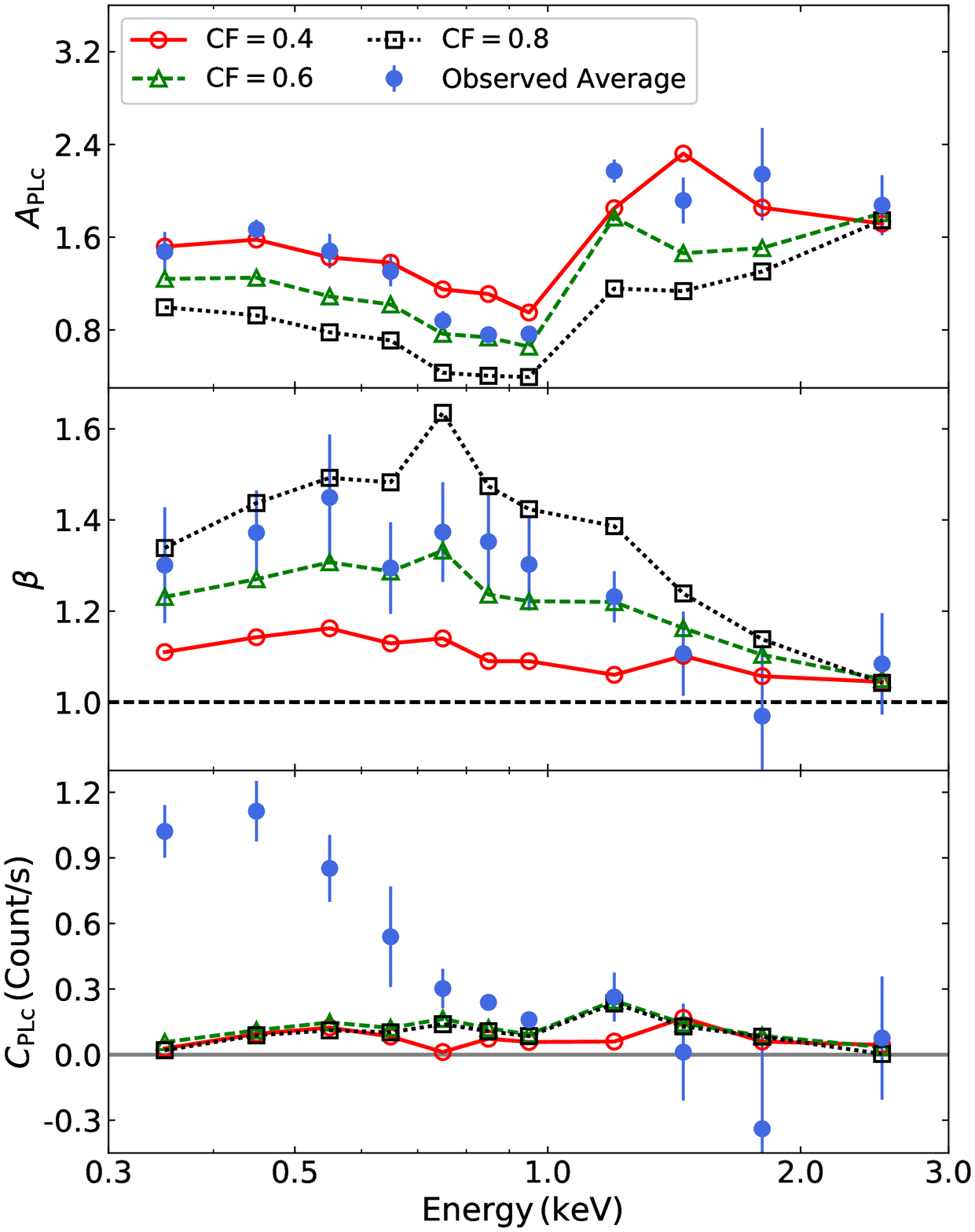}

\caption{The best-fit observed (filled symbols) and simulated (open symbols) 
parameters obtained by fitting the observed and simulated FFPs with a PLc model.}

\label{figapp:warmabs}
\end{figure}

To construct the model FFPs, we estimated the model count rate in the reference and the low-energy bands, assuming 10 different values of $N_{\rm PL}$ (between $N_{\rm PL,min}$ and $N_{\rm PL, max}$). Then we fitted them with a PLc model, exactly as we did with the observed FFPs. The best-fit simulated PLc parameters are plotted as empty symbols in Fig.\,\ref{figapp:warmabs}. 

In general, the assumed variable warm absorber model results in FFPs which are, qualitatively, similar to the observed plots. In all cases, $B_{\rm PLc,\, sim}$'s are steeper than one, as observed. Therefore, a variable warm absorber can produce non-linear FFPs, with slopes steeper than one. In addition, variable warm absorption can also result in non-zero, positive constants. But, the value of  $C_{\rm PLc,\, sim}$, at all energies, below 1\,keV is quite smaller than $C_{\rm PLc,\, obs}$. We also tried different $N_{\rm H}$ and/or CF values, and we saw that in some cases, a variable warm absorber model may even result in negative $C_{\rm PLc,\, sim}$ in the FFPs. In this case, the amplitude of the intrinsic constant spectral component will be larger than what $C_{\rm PLc,\, obs}$'s imply. 

\bsp	
\label{lastpage}
\end{document}